\documentclass[aps,pre,twocolumn,superscriptaddress,showpacs,showkeys,amsmath,amssymb]{revtex4-1}

\usepackage{epsfig,amsmath,amssymb,color}
\usepackage[T1]{fontenc}

\def\ket#1{\, | \, {#1} \, \rangle}

\begin{document}

\title{Electric field driven flat bands: 
Enhanced magnetoelectric and electrocaloric effects in frustrated
quantum magnets}

\author{Johannes Richter}
\affiliation{Institut f\"{u}r Physik, Universit\"{a}t Magdeburg, P.O. Box 4120, D-39016 Magdeburg, Germany}
\affiliation{Max-Planck-Institut f\"{u}r Physik Komplexer Systeme,
N\"{o}thnitzer Stra{\ss}e 38, D-01187 Dresden, Germany}

\author{Vadim Ohanyan}
\affiliation{Max-Planck-Institut f\"{u}r Physik Komplexer Systeme,
N\"{o}thnitzer Stra{\ss}e 38, D-01187 Dresden, Germany}
 \affiliation{Laboratory of Theoretical Physics, and Joint Laboratory of Theoretical Physics -- ICTP Affiliated Centre in Armenia,
          Yerevan State University,
         1 Alex Manoogian Str., 0025 Yerevan, Armenia}
\affiliation{CANDLE, Synchrotron Radiation institute, 31 Acharyan Str., 0040 Yerevan,
Armenia}

\author{J\"org Schulenburg}
\affiliation{Universit\"atsrechenzentrum, Universit\"at Magdeburg, D-39016
Magdeburg, Germany}

\author{J\"urgen Schnack}
\affiliation{Fakult\"at f\"ur Physik, Universit\"at Bielefeld, Postfach
100131, D-33501 Bielefeld, Germany}

\date{\today}

\begin{abstract}
The $J_1$-$J_2$ quantum spin sawtooth chain is a paradigmatic one-dimensional 
frustrated quantum spin system exhibiting unconventional ground-state and
finite-temperature properties.
In particular, it exhibits a flat energy band of one-magnon
excitations accompanied by
an enhanced magnetocaloric effect for two singular ratios
of the basal interactions $J_1$ and the zigzag interactions $J_2$.
In our paper, we demonstrate that one can drive the spin system into a
flat-band scenario by applying an appropriate electric field,
thus overcoming the restriction of fine-tuned exchange couplings
$J_1$ and $J_2$ and allowing one to tune more materials towards flat-band
physics, that is to show a macroscopic magnetization jump when
crossing the magnetic saturation field, a residual entropy at zero
temperature as  well as an enhanced  magnetocaloric effect.
While the magnetic field acts on the spin system via the
ordinary Zeeman term, the  coupling of an applied electric field
with the spins is given by the sophisticated Katsura-Nagaosa-Balatsky (KNB) mechanism, 
where the electric field effectively acts as a  Dzyaloshinskii-Moriya
spin-spin interaction. 
The resulting novel features are corresponding reciprocal
effects:
We find a magnetization jump driven by the electric field as well 
as a jump of the electric polarization driven by the
magnetic field, i.e.\ the system exhibits an extraordinarily strong
magnetoelectric effect.
Moreover, in analogy to the enhanced magnetocaloric effect the
system shows an enhanced electrocaloric effect. 

\end{abstract}



\maketitle

\section{Introduction.}
\label{sec2}

The magnetoelectric effect (MEE) allows to manipulate magnetic
materials by electric fields \cite{mee1}. Such an approach
promises several fundamental advantages since electric fields can be
manipulated on shorter time scales and can be confined to
smaller regions compared to magnetic fields. To drive future
quantum devices by means of electric control is thus at the
focus of substantial research activities in fields such as
energy transformation, sensors, magnetic storage, and spintronics
\cite{Ederer2008,Spaldin2008,GBB:S10,dong15,mee1,wang2014,mee2,fie16,spa19,don19,app1,app2,MEE_PRL2020,LLI:SA21}. 

Related to the aspect of energy conversion is the electrocaloric
effect (ECE), i.e.\ the ability to change temperature by
changing the electric field similar to the more familiar
magnetocaloric effect (MCE) \cite{ECE-review}. The renewed
interest is mainly stimulated by materials research on
ferroelectric thin films showing a strongly enhanced
ECE~\cite{science2006,science2008,nano2019} which opens the
window for future solid-state refrigeration technologies based
on the ECE \cite{science2017,science2020}.

Quantum systems hosting flat bands in their energy spectrum, on
the other hand, constitute realizations of materials that
already exhibit enhanced magnetocaloric effects thanks to the
special frustrated nature of their interactions. These systems appear in
different branches of physics such as highly frustrated
magnetism, strongly correlated electronic systems, cold atoms in
optical lattices, photonic lattices as well as twisted graphene
bilayers \cite{Tsui1982,25,26,27,2,spin-peierls,lm2,huber2010, 
kastura2010,kastura2011,JGT:PRL12,SOW:PRL12,bergholtz2013,sondhi2013,leykam2013,VCM:PRL15,MSC:PRL15,lm6,leykam2018,graphene2018,
photonic,mag_cryst_exp,prl2020,graphene2020a,graphene2020b,graphene2020c}.
Not only the enhanced MCE, but many intriguing phenomena such as
macroscopic magnetization jumps \cite{2,lm2,lm6} or fractional
quantum Hall physics \cite{Tsui1982,sondhi2013} are related to flat
bands. 

In the present paper we bring together flat-band phenomena and
magnetoelectric coupling. In particular, we demonstrate that one
can drive systems into the flat-band scenario by means of
electric fields and that one can thus achieve novel phenomena
such as a magnetization jump driven by the electric field as well 
as a jump of the electric polarization driven by the
magnetic field, and in analogy to the enhanced magnetocaloric
effect we find an enhanced electrocaloric effect.

   \begin{figure}[h]
 \begin{center}
\includegraphics[width=75.5mm]{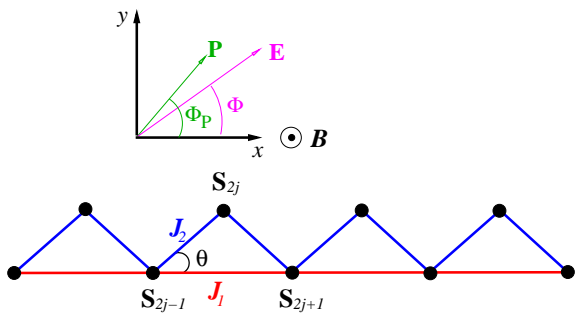}
\caption{(Color online) Sketch of the sawtooth chain together with the electric field 
$\mathbf{E}$, the $z$-aligned magnetic field $\mathbf{B}=B\mathbf{e}_z$ and the resulting electric polarization
$\mathbf{P}$. For the plane defined by 
the sawtooth geometry the $x$-$y$ plane is chosen.
 }
\label{fig1}
\end{center}
\end{figure}

   The MEE in most common  terms can be described as the magnetic-field
dependence of dielectric polarization and vice versa the electric-field
dependence of the magnetization in solids. 
The origin of the coupling between spins and the dielectric polarization can be very different \cite{mee1,mee2,dong15,don19}.
\textcolor{black}{ The one to be considered in the present paper is based on the so-called spin-current model or inverse Dzyaloshinskii-Moriya (DM) model and is called the
Katsura-Nagaosa-Balatsky (KNB) mechanism \cite{KNB1,KNB2,KNB3}.
According to the KNB mechanism an electric field can induce a DM term. This term
can be obtained 
by a second-order perturbation theory starting from a Hubbard-like model with
spin-orbit coupling.  
The electric field leads to a modification of the hopping terms
and to the emergence of new hopping paths. 
As a result, the  KNB mechanism  links 
the dielectric polarization corresponding to a pair of spins at adjacent lattice sites with the following
expression:}
\begin{eqnarray}
\label{KNB}
\mathbf{P}_{ij}=\gamma_{ij} \;\mathbf{e}_{ij}\times\mathbf{S}_i\times\mathbf{S}_j,
\end{eqnarray}
where $\mathbf{e}_{ij}$ is the unit vector pointing from site $i$ to site $j$ and
$\gamma_{ij}$ is a microscopic
parameter characterizing the quantum chemical features of the bond between two ions  with
spins $\mathbf{S}_i$ and $\mathbf{S}_j$ \cite{KNB1,KNB3,KNB2}. 
\textcolor{black}{Over the past 15 years 
numerous theoretical studies on quantum spin models with KNB mechanism have
been published, see, e.g., 
\cite{bro13,Berakdar2014,Berakdar2016,Berakdar2017,thakur18,XYZ,White2020,mench15,sznajd18,sznajd19,baran18,oles,stre20,oha20}.
Interestingly, for various one-dimensional unfrustrated models with the KNB-mechanism an
exact solution is possible, see 
 \cite{bro13,XYZ,oha20,mench15,baran18,oles}. 
}

In recent years, there is a growing interest in the realization of the 
magnetoelectric effect
in low-dimensional magnetic compounds.
A prominent class of these (quasi-)one-dimensional materials is given by edge-shared
cuprates with ferromagnetic (FM) nearest-neighbor ($J_1<0$) and antiferromagnetic
(AFM)  next-nearest neighbor ($J_2>0$)
interactions between the Cu$^{2+}$ ions carrying spin $1/2$, see, e.g., Refs.
\cite{mee2,LiCuO21,LiCu2O21,LiCu2O22,LiCuVO44,LiVCuO41,LiVCuO42,LiCuO23,Enderle2005,KBK:PRB06,Drechsler2007,LiCuVO43}.

At the same time,  many different frustrated
spin-lattice models in dimension $D=1,2,3$ were found,
where the lowest band of one-magnon excitations
above the FM vacuum is dispersionless (flat)
\cite{lm5,lm5a}.
Caused  by the flat band a number of unconventional features emerge
in  magnetic fields, such as a magnetization jump at the
saturation field $B_{\rm sat}$ \cite{2,lm4}, a magnetic-field driven
spin-Peierls instability \cite{spin-peierls},
magnon-crystallization in $D=2$ \cite{lm2,lm2a,prl2020},
a
finite residual entropy at $B_{\rm sat}$
\cite{lm2,lm3,lm2a,lm5,lm5a},
a very strong magnetocaloric effect
\cite{lm2a,cool2004,lm5}, and an additional
low-temperature maximum of the
specific heat signaling the occurrence of an additional
low-energy scale \cite{lm5,lm5a}.
Below we will demonstrate that due to the magnetoelectric coupling there are electric
analogues to the
above mentioned magnetization
jump and the  enhanced magnetocaloric effect.

A  prototype flat-band model
is the sawtooth (or delta) chain that has been widely
investigated for different realizations such as frustrated quantum spin systems, see, e.g.\
\cite{2,lm2,lm3,lm4,lm5,sawt2008,FAF1,lm8,lm9,FAF2,dre20,McClarty2020,Pujol2020,Hatsugai2020},
electronic systems, see, e.g.,
Refs.~\cite{25,26,27,28,29,hubb2010,31,32,maimaiti2017,33,Katsura_delta1,Katsura_delta2}
as well as photonic lattices, see, e.g.\ Refs.~\cite{photonic,34}.

In the present paper we consider the Heisenberg spin-half
sawtooth chain coupled to a $z$-aligned magnetic field $\mathbf{B}$ (Zeeman term) and to an electric
field $\mathbf{E}$  (KNB term)  with arbitrary direction but located within the plane defined by
the sawtooth geometry.   
The corresponding Hamiltonian is given by 
   \begin{eqnarray}\label{Ham}
   \mathcal{H}&=&J_1\sum_{j=1}^{N/2} \mathbf{S}_{2j-1}\cdot\mathbf{S}_{2j+1}+J_2\sum_{j=1}^{N/2}\mathbf{S}_{2j}\cdot\left(\mathbf{S}_{2j-1}+\mathbf{S}_{2j+1}\right)\nonumber\\
   &&-BM-\mathbf{E}\cdot\mathbf{P} \; , \;\mathbf{B}=B\mathbf{e}_z\; , \;M=S_z=\sum_{j=1}^N S_{j}^z , \\ 
   \mathbf{P}&=&
   \sum_{j=1}^{N/2}\mathbf{P}_{2j-1,2j+1}+\sum_{j=1}^{N/2}\left
   (\mathbf{P}_{2j-1,2j}
   +\mathbf{P}_{2j,2j+1}
   \right ) ,
   \end{eqnarray}
see Fig.~\ref{fig1} for the arrangement of spins and bonds as well as the  electric
and magnetic fields.
While we consider AFM $J_1$, the zigzag bond $J_2$ can be AFM or FM,
 however, restricted  to the
 region $2J_1 \le J_2 \le -2J_2$, where flat-band physics is possible.
  In case of zero electric field, for this model two flat-band scenarios are known:
For the AFM sawtooth chain, $J_1,J_2>0$, the lowest band of one-magnon excitations above the fully
polarized FM state $|\rm FM \rangle =|\uparrow\uparrow\uparrow
\ldots \rangle$ is  
dispersionless  for $J_2=2J_1$ (flat-band point) \cite{lm2, lm3, lm4, lm5, lm6, lm8,lm9}. 
As a result there exist localized multi-magnon
states (also called flat-band states)   for $N/4 \leq S_z < N/2$, which are the lowest states in the respective sectors
of $S_z$. At half of the saturation magnetization, $S_z = N/4$, there is a
wide plateau and the plateau state is a magnon-crystal state  \cite{2}.   
All the flat-band states are linearly independent \cite{lin_indep} and their number growths exponentially with the  number of sites $N$.
In magnetic fields close  to saturation 
the flat-band states dominate the low-temperature thermodynamics of the
model \cite{lm2, lm3, lm4, lm5, lm6, lm8,lm9}.
The experimental observation is not straightforward because the relevant
physics typically takes place at (very) high magnetic fields.
For a possible experimental realization of an alternative flat-band spin
system, where the saturation field is accessible, see
\cite{Tanaka2014,bil2018}.

  The second  flat-band scenario
is realized for 
the sawtooth Heisenberg  chain with FM bonds between the apical and basal spins ($J_2<0$) and AFM
bonds ($J_1>0$) within the basal line \cite{FAF1,FAF4,FAF5,FAF3,FAF_field}.
Here, the flat-band point  where the lowest band of one-magnon excitations above the fully
polarized FM state is dispersionless  is given by  $J_2=-2J_1$  \cite{FAF1}. 
Notably, the  flat-band states  exist here in all magnetization sectors 
$0 \leq S_z < N/2$, and again they are the lowest states in the respective
magnetization sectors.
The number of flat-band states growths
even faster exponentially with $N$ than for the first (purely
AFM) flat-band scenario leading again to the dominance of these
states at low $T$ \cite{FAF1,FAF3,lm9}.    
A striking difference to scenario 1
is that   
flat-band physics takes place at zero-magnetic field which enhances the
chance to observe the flat-band physics in an experiment.
Indeed, there are magnetic compounds which are well described by the
FM-AFM sawtooth spin
chain, such as   malonate-bridged copper complexes
\cite{FAF7,FAF8} and the very recently  synthesized and studied magnetic molecule
Fe$_{10}$Gd$_{10}$ \cite{FeGd}. While the parameter situation in the former one is not very close to
the flat-band point,
the 
Fe$_{10}$Gd$_{10}$   molecule exhibits exchange parameters close to the
flat-band point \cite{FeGd}, and, therefore signatures of flat-band physics
are well seen in this system \cite{FeGd,FAF3,FAF_field}.

Further examples for magnetic compounds with sawtooth chain geometry of
exchange bonds are the
atacamite Cu$_2$Cl(OH)$_3$ \cite{atacamite},
the fluoride Cs$_2$LiTi$_3$F$_{12}$ \cite{fluoride}, the euchroite
Cu$_2$(AsO$_4$)(OH)$\cdot$3H$_2$O  \cite{euchroite}, the  sawtooth spin ring
Mo$_{75}$V$_{20}$ \cite{Mo75V20},
\textcolor{black}{ 
the frustrated [Mn$_{18}$] magnetic wheel-of-wheels molecule \cite{Mn18} and
the iron compound  Fe$_2$Se$_2$O$_7$ \cite{sawtooth_exp_2021}.}

\subsection{Summary of results}
The key target
of the present paper is the study of the interplay of the KNB mechanism and
magnetic frustration at and in the vicinity of a flat-band point.
As mentioned above, for the model at hand  at zero electric
field the realization of
flat-band physics 
requires fine tuning of the exchange parameters, i.e., the very existence of a
strictly  flat one-magnon band appears only at two singular ratios $J_2/J_1$.
Naively, one may expect that the presence of the electric field leading to
additional DM terms
in the Hamiltonian 
 eliminates  flat-band phenomena.
\begin{itemize}
\item
A crucial finding of our work is that actually   
just the electric field via the KNB mechanism
may dissolve the fine tuning of $J_1$ and $J_2$ and can lead to a large variety
 of $J_1$-$J_2$ ratios, where for appropriate direction and magnitude of
 $\mathbf{E}$ the lowest one-magnon band is flat.
\item 
Thus, for a certain system at hand with given values of $J_1$ and
 $J_2$ we can achieve flat-band physics by application of
an appropriate value of the electric field $E_{\rm f}(J_1,J_2)$ (flat-band field).
\item Moreover, the saturation field $B_{\rm sat}$ in the vicinity of which 
the flat-band physics can be observed is lower in the presence of an
electric field  than for the previously
studied flat-band situation at zero electric field
\cite{2,lm2,lm5,lm5a}, this way leading 
to  a better  access to flat-band physics in experiments.     
\item
The flat-band effects known from previous studies \cite{2,lm2,lm5,lm5a} without
electric field, such as a macroscopic
magnetization jump at saturation field,  the huge degeneracy of the ground states at the flat-band
point leading to a residual entropy, the emergence of an extra-low energy scale
in the vicinity   the flat-band
point as well as an  
enhanced magnetocaloric effect are also  present in case of the electric-field
driven flat-band physics.  
\item
In addition to these flat-band phenomena,
the presence of an electric  field  leads to intriguing reciprocal effects: 
macroscopic  jumps in the electric polarization $P$ driven  by  magnetic
field $B$ and macroscopic  jumps in the magnetization  $M$ driven 
by the electric field  $E$, i.e.,
there is an extraordinary  MEE.
\item
Last but not least, besides 
an enhanced magnetocaloric effect known for flat-band magnets 
there is an enhanced
electrocaloric effect. If considering adiabatic cooling for isentropes with
entropy $s$ below the
residual
entropy $s_{\rm res} \sim 0.24$, the temperature drops quickly to zero when
approaching  $E=E_{\rm f}(J_1,J_2)$.

\end{itemize}

\section{KNB mechanism for the sawtooth chain}

In this section we specify the KNB mechanism for the sawtooth-chain
geometry.
We choose the $x$ axis along the basal line of $J_1$ bonds and
the $x$-$y$ plane for
the location of the $J_1$-$J_2$-$J_2$ triangle,
where the angle between $J_1$ and $J_2$ bonds is $\theta$,  
see Fig.~\ref{fig1}. 
The unit vectors along the exchange bonds entering Eq.~(\ref{KNB}) read
\begin{eqnarray}
&&\mathbf{e}_{2j-1,2j+1}=\mathbf{e}_x,\\
&&\mathbf{e}_{2j-1,2j}=\cos\theta\; \mathbf{e}_x+\sin\theta\; \mathbf{e}_y,\nonumber\\
&&\mathbf{e}_{2j,2j+1}=\cos\theta\; \mathbf{e}_x-\sin\theta\; \mathbf{e}_y.\nonumber
\end{eqnarray}
Then the corresponding local polarization vectors are given by
\begin{eqnarray}\label{P_dc}
&&\mathbf{P}_{2j-1,2j}=\gamma\left(\cos\theta \mathbf{e}_x+\sin\theta \mathbf{e}_y\right)\times\mathbf{S}_{2j-1}\times\mathbf{S}_{2j}\nonumber ,\\
&&\mathbf{P}_{2j,2j+1}=\gamma\left(\cos\theta\mathbf{e}_x-\sin\theta\mathbf{e}_y\right)\times\mathbf{S}_{2j}\times\mathbf{S}_{2j+1}\nonumber ,\\
&&\mathbf{P}_{2j-1,2j+1}=\gamma^{\prime}\mathbf{e}_x\times\mathbf{S}_{2j-1}\times\mathbf{S}_{2j+1},
\end{eqnarray}
where a different KNB parameter $\gamma^{\prime}=\alpha\gamma$ is considered
for the polarization in the basal line
taking care of the possibility to have different microscopic quantum-chemical parameters of the KNB
mechanism for different exchange bonds.
We can merge the first two relations of the Eqs.~(\ref{P_dc})
by rewriting them appropriately 
in one equation: 
\begin{equation}\label{P_dc_2}
\mathbf{P}_{j,j+1}=\gamma \left(\cos\theta\;\mathbf{e}_x-(-1)^j\sin\theta \;\mathbf{e}_y\right)\times\mathbf{S}_{j}\times\mathbf{S}_{j+1}.
\quad
\end{equation}
Note that this expression coincides with the one considered for
the $XY$ zigzag chain in
Ref.~\cite{baran18}.
For an electric field $\mathbf{E}=(E_x,E_y,0)$ residing in the
$x$-$y$ plane, see Fig.~\ref{fig1},
the interaction between the polarization and the electric field entering
the Hamiltonian (\ref{Ham})
reads
\begin{eqnarray}\label{EP}
&-&\mathbf{E}\cdot\mathbf{P}\\
 &=&\sum_{j=1}^N\left(\gamma E_y\cos\theta+(-1)^j\gamma E_x\sin\theta\right)\left(S_j^xS_{j+1}^y-S_j^yS_{j+1}^x\right)\nonumber \\
 &+&\alpha\gamma E_ y
 \sum_{j=1}^{N/2}\left(S_{2j-1}^xS_{2j+1}^y-S_{2j-1}^yS_{2j+1}^x\right). \nonumber
\end{eqnarray}
Note that for this specific combination of the $x$- and $y$-components
of the spin operator $\mathbf{S}_i$  the Hamiltonian
(\ref{Ham}) commutes with the $z$-component of the total spin, $S_z$.

For convenience, 
we absorb the KNB constant $\gamma$ in $\mathbf{E}$, which in turn is measured in appropriate units.
The observables relevant for the MEE  are the 
expectation values of the $x$ and $y$ components of total polarization and
the $z$-aligned magnetization:
\begin{eqnarray}
P_x&=&\frac {\sin\theta\langle\sum_{j=1}^{N}(-1)^j\left(S_j^yS_{j+1}^x-S_j^xS_{j+1}^y\right)\rangle}{N},\nonumber \\
P_y&=&\frac {\cos\theta\langle\sum_{j=1}^{N}\left(S_j^yS_{j+1}^x-S_j^xS_{j+1}^y\right)\rangle}{N},\nonumber \\
&&-\alpha\;\frac{\langle\sum_{j=1}^{N/2}\left(S_{2j-1}^xS_{2j+1}^y-S_{2j-1}^yS_{2j+1}^x\right)\rangle}{N},\nonumber \\
M&=&M_z=\frac{\langle\sum_{j=1}^{N}S_j^z\rangle}{N},
\end{eqnarray}
where $\langle\cdot\rangle$ denotes either the expectation value with
respect to a specific state or the thermal average.

Let us mention here, that the considered interaction term (\ref{EP}) can be
also
understood as DM interaction
$\mathbf{D}_i=D_i\mathbf{e}_z$, $i=1,2,3$, with 
\begin{eqnarray} \label{D_i}
&&D_1=\alpha E \sin\phi, \\ 
&&D_2=E(\sin\phi\cos\theta-\cos\phi\sin\theta), \nonumber \\
&&D_3=E(\sin\phi\cos\theta+\cos\phi\sin\theta), \nonumber  
\end{eqnarray}
where $D_1$ belongs to the basal bonds, and $D_2$  and $D_3$ belong to the two
zigzag bonds, respectively.
Such a DM term could be relevant for spin lattices with
low symmetry. 

\section{Electric field induced flat bands}
In this section we will figure out how the lowest one-magnon band can be flat
also in the presence of an  electric field.
We start with the fully polarized FM state $|FM\rangle=|\uparrow\uparrow\ldots\uparrow\rangle$          
which is the ground state for strong enough magnetic field $B$.
Imposing 
periodic boundary conditions we construct one-magnon excitations above the
magnon vacuum $|FM\rangle$
\begin{eqnarray}
|1_k\rangle=\sum_{l=0,1}a_l\sum_{j=1}^{N/2}e^{i\;jk}S_{2j+l}^{-}|FM\rangle,
\end{eqnarray}
where $k$ is the quasi-momentum.
The calculation of the two branches (according to the two sites per unit
cell) of the one-magnon spectrum is straightforward:
\begin{widetext}
\begin{eqnarray}\label{1m_sp_gen}
&&\varepsilon^{\pm}(k)=B-\frac{J_1+2J_2}{2} +\frac{1}{2}\left[\widetilde{J}_1\cos(k-k_1)\pm\sqrt{\left(J_1-\widetilde{J}_1\cos(k-k_1)\right)^2
+2\widetilde{J}_2\widetilde{J}_3\cos(k-k_2-k_3)+\widetilde{J}_2^2+\widetilde{J}_3^2}\right]\;\\
&& \widetilde{J}_1=\sqrt{J_1^2+D_1^2}, \widetilde{J}_2=\sqrt{J_2^2+D_2^2},
\widetilde{J}_3=\sqrt{J_2^2+D_3^2}, k_1=\arctan\frac{D_1}{J_1},
k_2=\arctan\frac{D_2}{J_2},
k_3=\arctan\frac{D_3}{J_2}. \label{tylde_J}
\end{eqnarray}
\end{widetext}
Because the expression for the spectrum (\ref{1m_sp_gen}) is a
bit cumbersome due to the phase shifts $k_i$,
we will consider two special cases for which simplified expressions for
$\varepsilon^{\pm}(k)$ are obtained that enable to extract criteria in
analytical form for the very existence of
flat-band physics.

\subsection{Flat-band case I} 
Here we consider  
an electric field pointing along
the $x$-axis, i.e., $\phi=0$. In this case
we have $D_1=0$, $D_2=-D_3=-E\sin\theta$ yielding for the phase shifts
$k_1=0$ and $k_2=-k_3$,  and the one-magnon dispersion then 
reads
\begin{eqnarray}\label{1-magn1}
&&\varepsilon^{\pm}(k)=B-\frac{J_1+2J_2}{2}+\frac 12 \Big (J_1\cos k \\
&&\pm\sqrt{J_1^2(1-\cos k)^2+2(J_2^2+E^2\sin^2\theta)(1+\cos k)} \Big ). \nonumber
\end{eqnarray}
Obviously, this expression is very similar to the known standard cases,
namely replacing  $J_2^2+E^2\sin^2\theta$ by an effective coupling $J_{2,\text{eff}}^2$
we just obtain the expression for the $k$-dependent term for the pure Heisenberg
model without KNB terms, see, e.g., Eq.~(3) in
\cite{sawt2008}.
Thus, one gets a flat band 
if the  $x$-aligned  electric field obeys the relation   
\begin{eqnarray}\label{E_1}
E= E_{\rm f}=\pm \frac{\sqrt{4J_1^2-J_2^2}}{\sin\theta}.
\end{eqnarray}
This equation also implies that a flat band driven by an  
electric field does exist for arbitrary values of $J_1$ and
 $J_2$ with the constraint
$4J_1^2 \ge J_2^2$.

Inserting  (\ref{E_1})
in Eq.~(\ref{1-magn1}) we obtain:
\begin{eqnarray}\label{1-magn1-b}
\varepsilon^{+}(k)&=&B+J_1-J_2+J_1\cos k \nonumber \\ 
\varepsilon^{-}(k)&=&B-2J_1-J_2 .
\end{eqnarray}
This yields for the saturation (flat-band) field  
\begin{eqnarray}\label{Bsat}
B_{\rm f}=B_{\rm sat}=2J_1+J_2 ,
\end{eqnarray}
which is lower than 
the corresponding value  $B_{\rm sat}=4J_1$ for the standard AFM  flat-band
case. 
Let us mention that the above discussed flat-band scenario 
also corresponds to a purely magnetic system without electric field, i.e.,
for the $J_1$-$J_2$   
      Heisenberg sawtooth chain with a staggered DM-interaction term in $z$-direction along the zigzag
 $J_2$ bonds, $D\sum_{j=1}^N(-1)^j\left(S_j^xS_{j+1}^y-S_j^yS_{j+1}^x\right)$,
 if $D=\sqrt{4J_1^2-J_2^2}$.  

We conclude that the above outlined case I for the model at hand opens the
window for a
flexible access to flat-band physics via the KNB mechanism, because no fine-tuning of the exchange
parameters $J_1$ and $J_2$ is necessary.
Moreover, the reduced  saturation field also
improves the  possibility  to
have experimental access to flat-band physics.

\subsection{Flat-band case II}
Let us consider a second specific case allowing a simplification of 
Eq.~(\ref{1m_sp_gen}). It is given if the electric field is directed parallel to the
zigzag bonds, i.e., either $\phi=\pm\theta$ or $\phi=\pm(\pi-\theta)$.
Without loss of generality we will take the sign to be plus. 
The corresponding DM terms become  $D_1=\alpha E\sin\theta$, $D_2=0$,
and $D_3=2E\sin\theta\cos\theta$ which leads to
 $\widetilde{J}_2=J_2$ and $k_2=0$  in Eqs.~(\ref{1m_sp_gen})  and
 (\ref{tylde_J}).
A flat band is possible if the remaining phase shifts $k_1$ and $k_2$ in
Eq.~(\ref{1m_sp_gen}) are equal, i.e.,
\begin{eqnarray}\label{relJ1J2}
J_2=\frac{2\cos \theta}{\alpha}J_1.
\end{eqnarray}
Finally,   the lower band of the one-magnon excitations becomes flat 
for 
\begin{eqnarray}\label{E_flat2}
E=\pm \frac{2J_1\sqrt{\cos^2\theta-\alpha^2}}{\alpha^2\sin\theta} \; , \;
|\alpha| \leq |\cos \theta| .
\end{eqnarray}
Obviously, the possible values for $\alpha$ are restricted by
the bond angle $\theta$. 
As in the previous case the two signs of $E$ correspond to the symmetry $\theta\rightarrow -\theta$, $E\rightarrow
-E$. Note, however, that for $\phi=-\theta$, the DM-terms in the zigzag part of the chain will be non-zero for
odd zigzag-bonds, and we have $D_1=-\alpha E \sin\theta$,
$D_2=-2E\sin\theta\cos\theta$, and $D_3=0$.

 Again it is appropriate to mention that the above discussed flat-band scenario
also corresponds to a purely magnetic system without electric field, i.e.,
for the $J_1$-$J_2$
      Heisenberg sawtooth chain with specific DM terms, namely
uniform DM-interaction along the basal line and non-zero DM-terms only for the even bond on the zigzag
 line,
\begin{eqnarray}
&&D_a\sum_{j=1}^N(1\pm(-1)^j)\left(S_j^xS_{j+1}^y-S_j^yS_{j+1}^x\right) \\
&&+D_b\sum_{j=1}^{N/2}\left(S_{2j-1}^xS_{2j+1}^y-S_{2j-1}^yS_{2j+1}^x\right)
\;, \;
\frac{D_a}{D_b}=\frac{J_2}{J_1}.\nonumber
\end{eqnarray}
Then the flat band is realized for $D_a=\pm\sqrt{J_2^2-4J_1^2}$.

We may conclude that by contrast to case I the above outlined flat-band scenario
II is less promising
with respect to a possible experimental realization because the precondition 
of fine-tuning of the exchange
parameters $J_1$ and $J_2$ is not removed.

 \begin{figure}[hb!]
\includegraphics[clip=on,width=75mm]{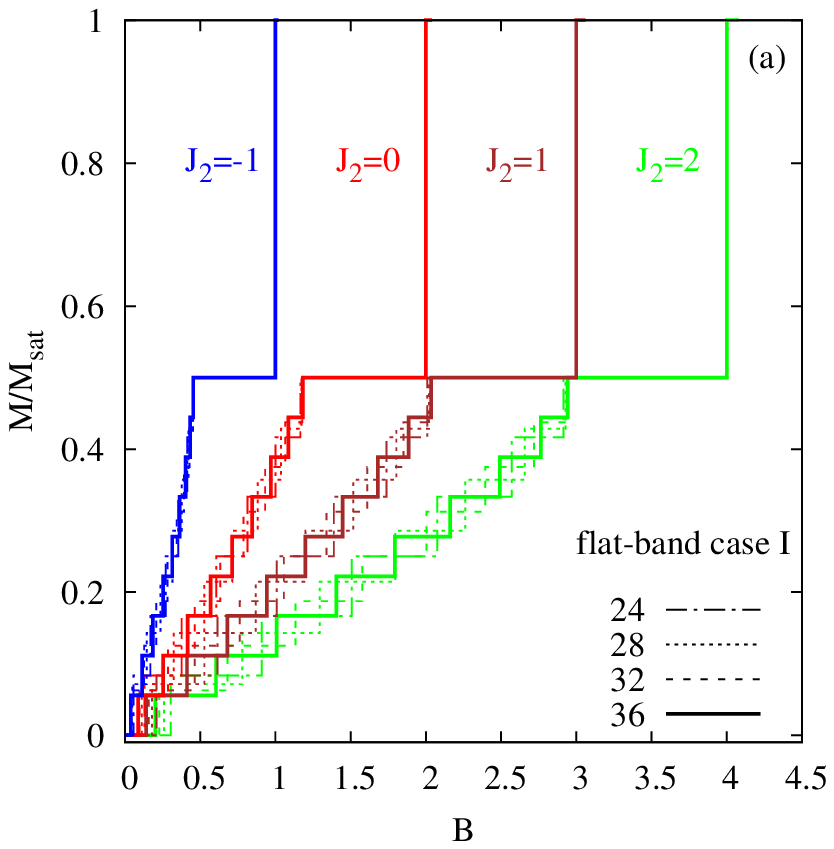}
\includegraphics[clip=on,width=75mm]{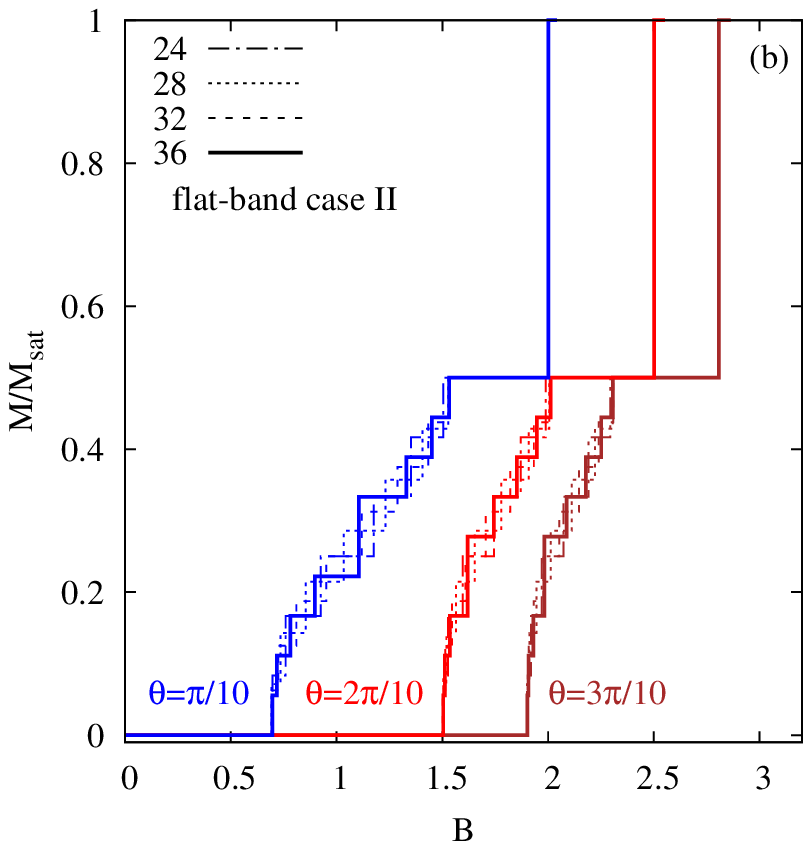}
\caption{(Color online)  GS data of the magnetization $M/M_{\rm sat}$
as a function of the magnetic field
$B$ for $N=24,28,32,36$.
(a) Flat-band case I (i.e., $\phi=0$) with $\theta=\pi/8$,
$J_1=1$ and various values of $J_2$.
(b) Flat-band case II (i.e., $\phi=\theta$) with $\alpha=1/2$ and
three values of $\theta$ (obeying $|\alpha| \leq |\cos \theta|$, see
Eq.~(\ref{E_flat2}))
and $\sqrt{J_1^2+J_2^2}=1$.
}
\label{fig2}
\end{figure}

\section{Numerical results}
We will focus here on case I as the promising case that does not require
fine-tuning of $J_1$ and $J_2$.
Case II will be discussed only secondarily.
Having in mind, that for case I the value of the bond angle enters the
Hamiltonian via $E\sin\theta$ only, cf. Eqs.~(\ref{D_i}) and (\ref{E_1}),   
we will show in what follows numerical data for one example of the bond angle,
namely 
$\theta=\pi/8$.

We use the exact diagonalization method (ED) to determine the ground state
(GS) as well as
finite-temperature properties for finite chains with periodic boundary
conditions. 
To exploit translational symmetry we perform the ED calculations using J\"org
Schulenburg's {\it spinpack}
code
\cite{49,50}.

The ED is a well established quantum many-body technique which is
widely applied to frustrated quantum spin systems. It is especially 
appropriate for one-dimensional systems, because one can apply the ED to
a set of finite systems with various numbers of spins $N$. Thus, by comparing
systems with different $N$ the finite-size effects can be controlled.
Moreover, from previous studies \cite{2,FAF1,Accuracy2020}
it is known that finite-size effects can be 
 particularly small for the sawtooth spin chain. 
In the present paper we calculate the GS up to $N=36$ by using the
Lanczos method
and the 
thermodynamics by calculating  the full spectrum  up to $N=20$.

In addition to the ED we apply 
the approximate finite-temperature Lanczos
method (FTLM) to calculate thermodynamic properties
of larger chains $N > 20$.  
FTLM is a Monte-Carlo like extension of the full ED.
Thermodynamic quantities are determined using trace estimators
\cite{51,52,53,54,55,56,57,58,59,61,62}.
The partition function $Z$ is approximated by a Monte-Carlo like representation of $Z$, 
i.e., the sum over a complete set of $(2s+1)^N$ basis states entering $Z$
is replaced by a much smaller sum over $R$ random vectors $\ket{\nu}$ for each subspace ${\mathcal H}(S_z)$ of the Hilbert
space.

 \begin{figure}[ht!]
 \begin{center}
\includegraphics[clip=on,width=75mm]{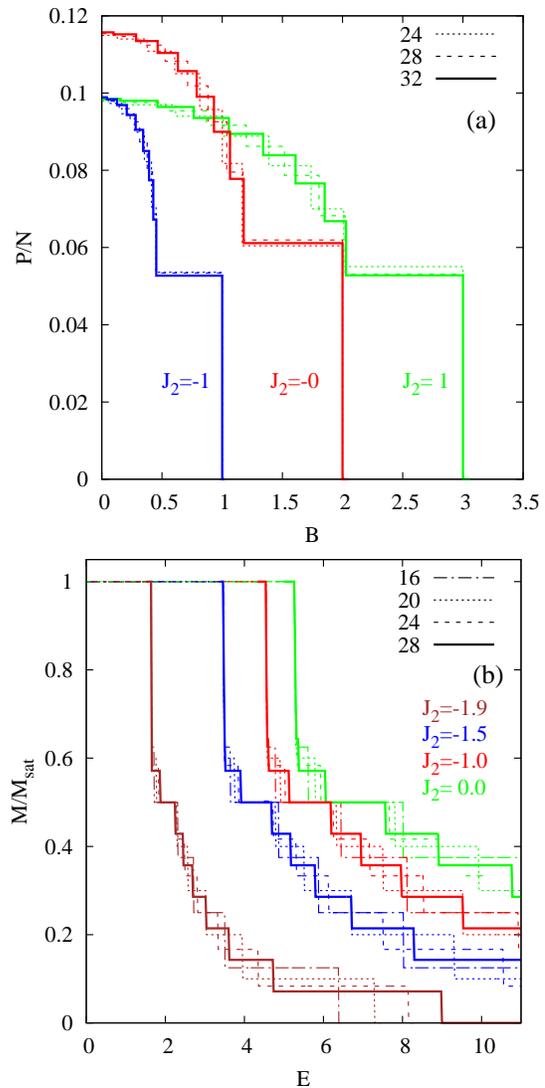}
\caption{(Color online)  MEE for flat-band case I (i.e., $\phi=0$)
with $\theta=\pi/8$,
$J_1=1$ and various values of $J_2$. The numbers in the legend denote the
system size $N$. 
(a) Electric polarization $P=\sqrt{P_x^2+P_y^2}$ as function of the magnetic field $B$ for  $E=E_{\rm f}$.      
(b) Magnetization  $M$ as function of the electric field $E$ for  $B=1.01B_{\rm sat}$. 
}
\label{fig3}
\end{center}
\end{figure}

\subsection{Ground-state properties}
We start with the flat-band case I and consider the magnetization curve
$M(B)$ first, see Fig.~\ref{fig2}(a).
$M(B)$ is typical for the known standard (i.e., $E=0$) AFM flat-band case
\cite{2,sawt2008}: There is a wide plateau at $M/M_{\rm sat}=1/2$ preceeding the jump
to
saturation. The jump is size-independent.
Moreover, the dependence of the width of the plateau on the system
size is very weak.
Only when approaching the standard FM-AFM flat-band case (i.e., $J_2 \to -2J_1$)
\cite{FAF1}
the plateau shrinks and finally the jump takes place directly from $M=0$ to
$M=M_{\rm sat}$.
Below the plateau the finite-size steps in $M(B)$ become 
smaller with growing $N$, and the magnetization curve gets
finally smooth 
without peculiarities as $N \to \infty$.

As for the standard AFM flat-band case there is a massively degenerated
 ground-state manifold at the saturation field which is built by the localized   
localized multi-magnon (flat-band) states leading to a residual entropy per site of 
$s_{\rm res}
=\frac{1}{2}\ln\frac{1+\sqrt{5}}{2}=0.240606$ \cite{lm2,lm3}.
Thus we conclude that   
the GS properties of the flat-band system driven by an  electric field $E$  are identical to   
those of the well-studied AFM flat-band system in absence of $E$.

We mention  here that for the flat-band case II the GS flat-band physics at
$B=B_{\rm sat}$ is identical
to case I, i.e., the magnetization jump and the magnon-crystal plateau 
are also
present, see Fig.~\ref{fig2}(b).
Moreover, the residual entropy at $B=B_{\rm sat}$ is the same for cases I and II.

 \begin{figure}[ht!]
 \begin{center}
\includegraphics[clip=on,width=75mm]{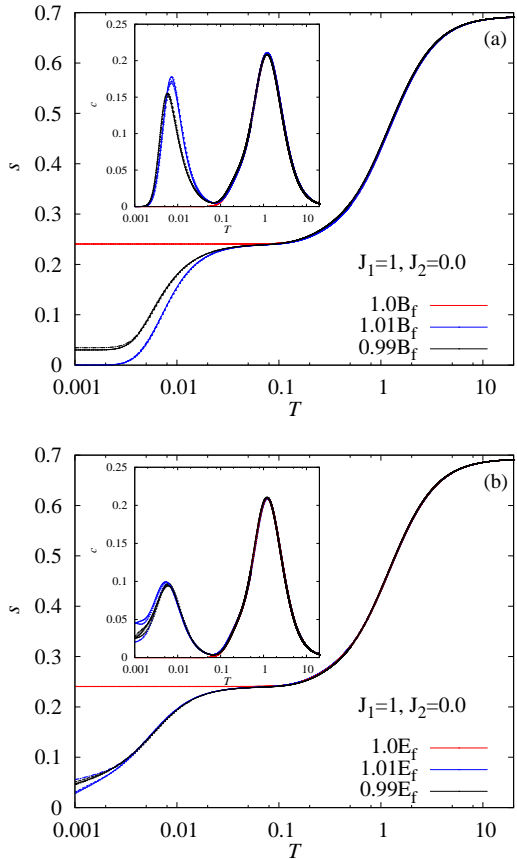}
\caption{(Color online)  Entropy $s=S/N$ (main panel) and specific heat
$c=C/N$ (inset) for the flat-band case I   
with $J_1=1$, $J_2=0$, $\phi=0$, and $\theta=\pi/8$ for $N=20$ (full ED,
short dashes),
$N=28$ (FTLM, long dashes) and $N=36$ (FTLM, solid) at and near the flat-band point 
$B=B_{\rm f}=B_{\rm sat}=2.0$
and  $E=E_{\rm f}=2/\sin(\theta)$.  
(a): $E=E_{\rm f}$, $B=1.01B_{\rm f}$, $1.0B_{\rm f}$, and $B=0.99B_{\rm f}$.  
(b)  $B=B_{\rm f}=2$, $E=1.01E_{\rm f}$, $1.0E_{\rm f}$, and
$E=0.99E_{\rm f}$.
Note, that the curves for different $N$ almost coincide.  
}
\label{fig4}
\end{center}
\end{figure}

\textcolor{black}{
The flat-band induced  size-independent discontinuous change of $M(B)$ upon crossing $B_{\rm f}$
leads to similar
abrupt changes of $P(B)$ when $B$ traverses the flat-band point $B_{\rm f}$. On the other hand, 
the electric field $E$ drives a jump-like behavior of the
magnetization $M(E)$.     
Since all these jumps are related to the very existence of a flat-band 
we may  expect that also the jumps in $P(B)$ and $M(E)$ 
  will exhibit at most only small finite-size effects.  }
We demonstrate this in Fig.~\ref{fig3}. In the upper panel (a) we show the
electric polarization $P$
as a function of the magnetic field $B$ for $J_1=1$, $\theta=\pi/8$  and various
$J_2$ values, where the electric field $E$ is set to the flat-band value
$E_{\rm f}$ given by 
 Eq.~(\ref{E_1}).   
Indeed, we observe a jump of $P$ at $B=B_{\rm sat}$ that amounts
to more than 50\% of the initial value $P$ at $B=0$.
As expected, finite-size effects are very small.   
 In the lower panel (b) we show the magnetization $M$
as a function of the electric field $E$ for $J_1=1$, $\theta=\pi/8$  and
 various $J_2$  values, where the magnetic field is set to $B=1.01B_{\text{sat}}$,
 i.e., for
values of the electric field below the  flat-band value $E_{\rm f}$  
 the GS is the fully polarized FM state. When $E$ crosses $E_{\rm f}$ 
we find a jump down to about 60\% of saturation. Interestingly, the  jump
is even larger for larger $N$. 
We may conclude that discontinuous changes in $M(E)$ and $P(H)$
most likely remain for $N \to \infty$.
Thus, we found evidence 
of an extraordinarily enhanced MEE with an abrupt change of $M$
(resp.\ $P$)
when varying the electric (resp.\ magnetic) field due to the very existence
of a flat band in the spin system at hand. This provides the
opportunity      
of switching the magnetization by an electric field or the electric
polarization by a magnetic field.

\begin{figure}[hbt!]
\begin{center}
\includegraphics[clip=on,width=75mm]{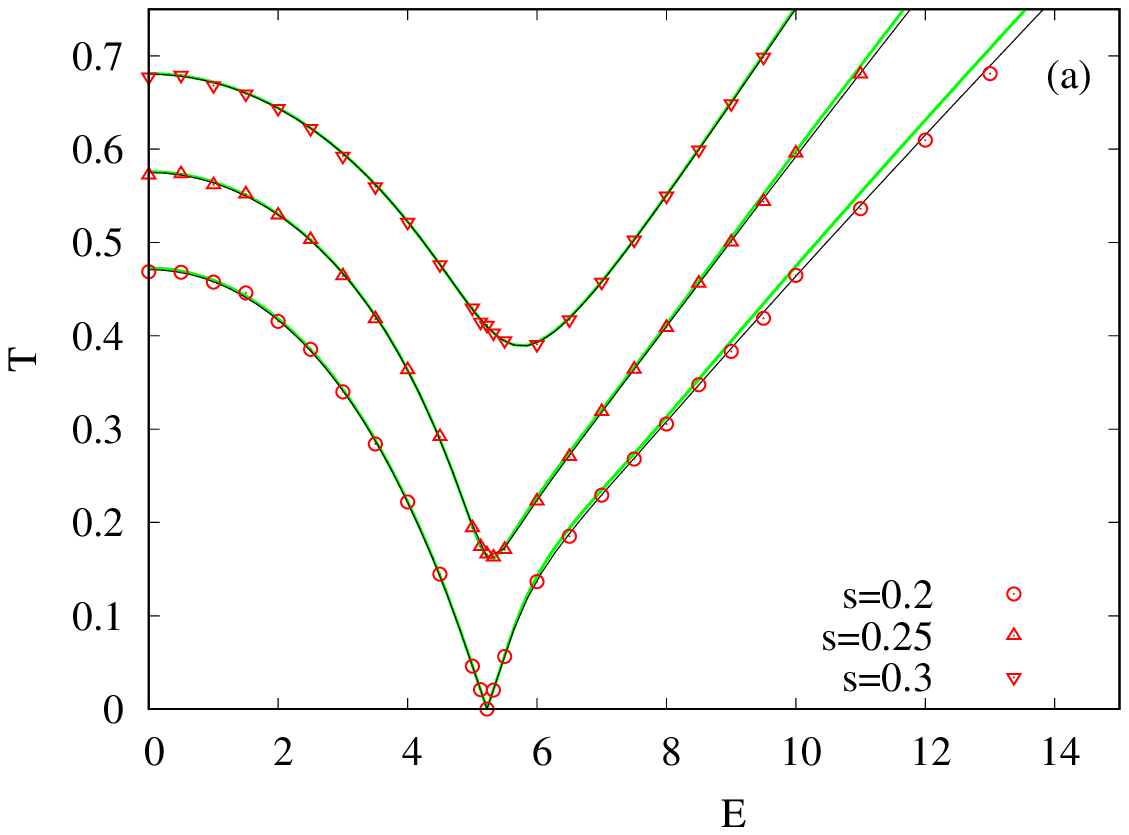}
\includegraphics[clip=on,width=75mm]{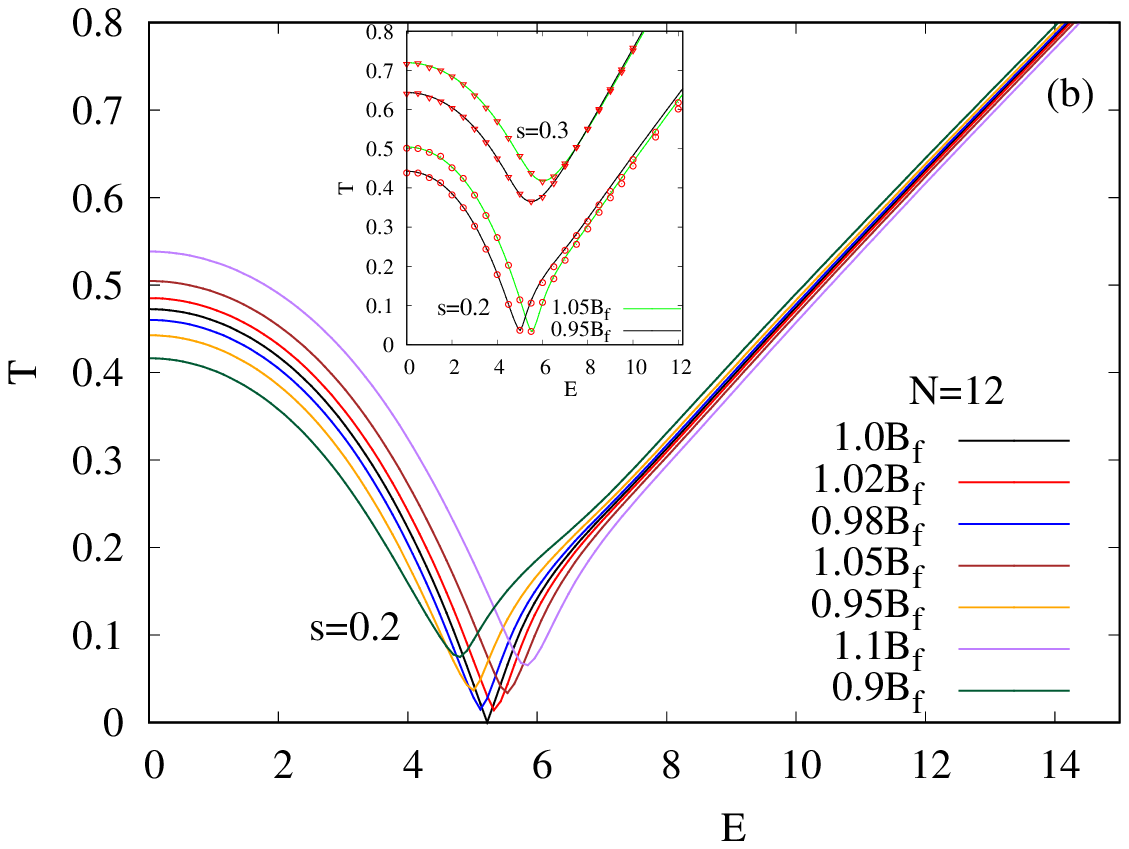}
\caption{(Color online)
\textcolor{black}{
(a)
Finite-size effects of the electrocaloric effect for the flat-band case I
($\phi=0$,
$\theta=\pi/8$, $B=B_{\rm f}$)
for $J_1=1$ and $J_2=0$ (green lines -- $N=12$, black lines -- $N=16$, red symbols
-- $N=24$).
Isentropes are shown for $s=S/N=0.2$, $0.25$,  and $0.3$.
(b) Main panel: The electrocaloric effect  (isentropes for $s=S/N=0.2$) for the flat-band case I
shown for magnetic field $B$ at various values below and above the flat-band value $B_{\rm f}$
for a chain of size $N=12$.
Inset: Finite-size effects of the electrocaloric effect shown for isentropes for $s=S/N=0.2$
and $0.3$ (lines -- $N=12$,  red symbols -- $N=24$).}
}
\label{fig5}
\end{center}
\end{figure}

\begin{figure}[ht!]
\begin{center}
\includegraphics[clip=on,angle=270,width=50mm]{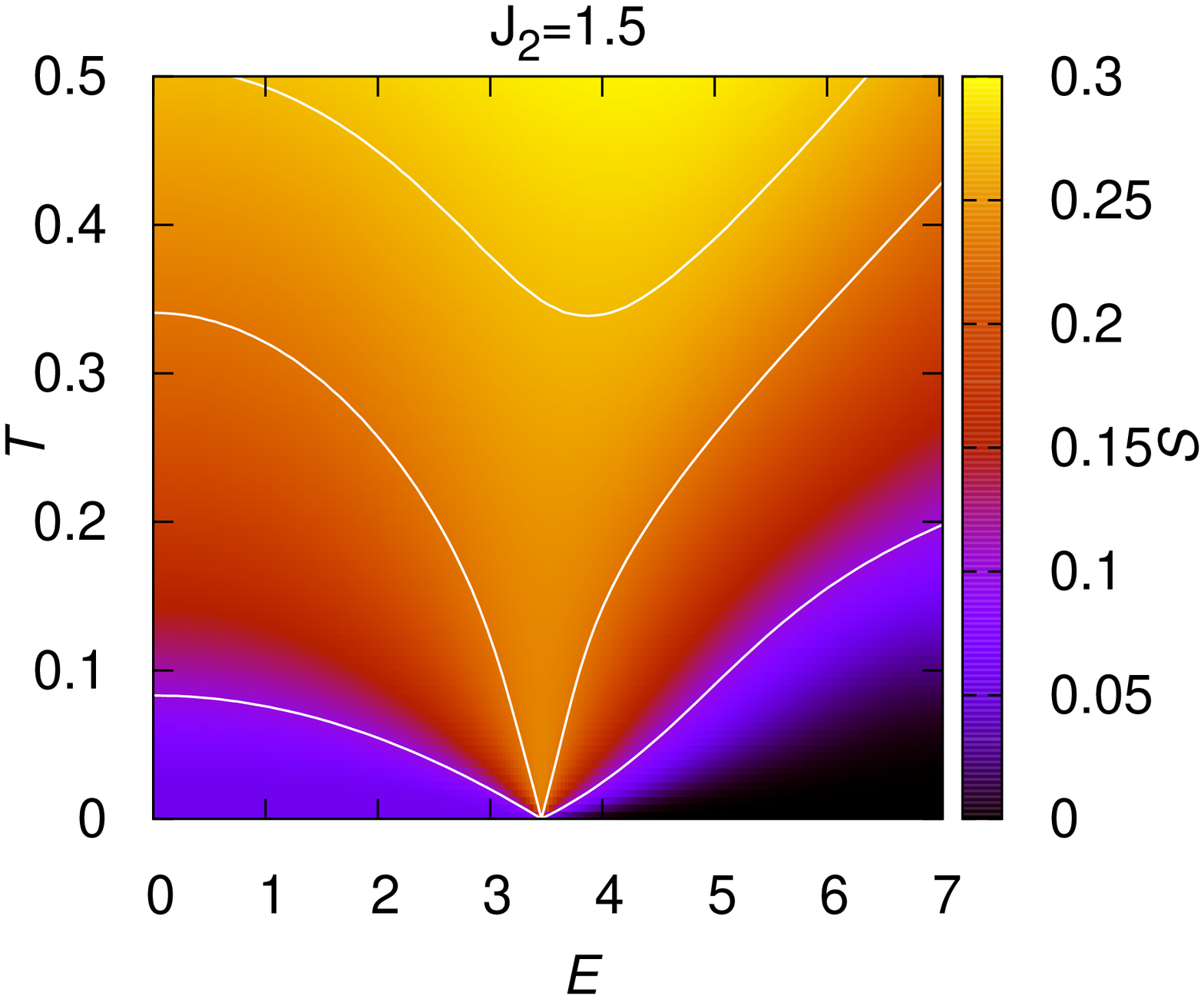}

\includegraphics[clip=on,angle=270,width=50mm]{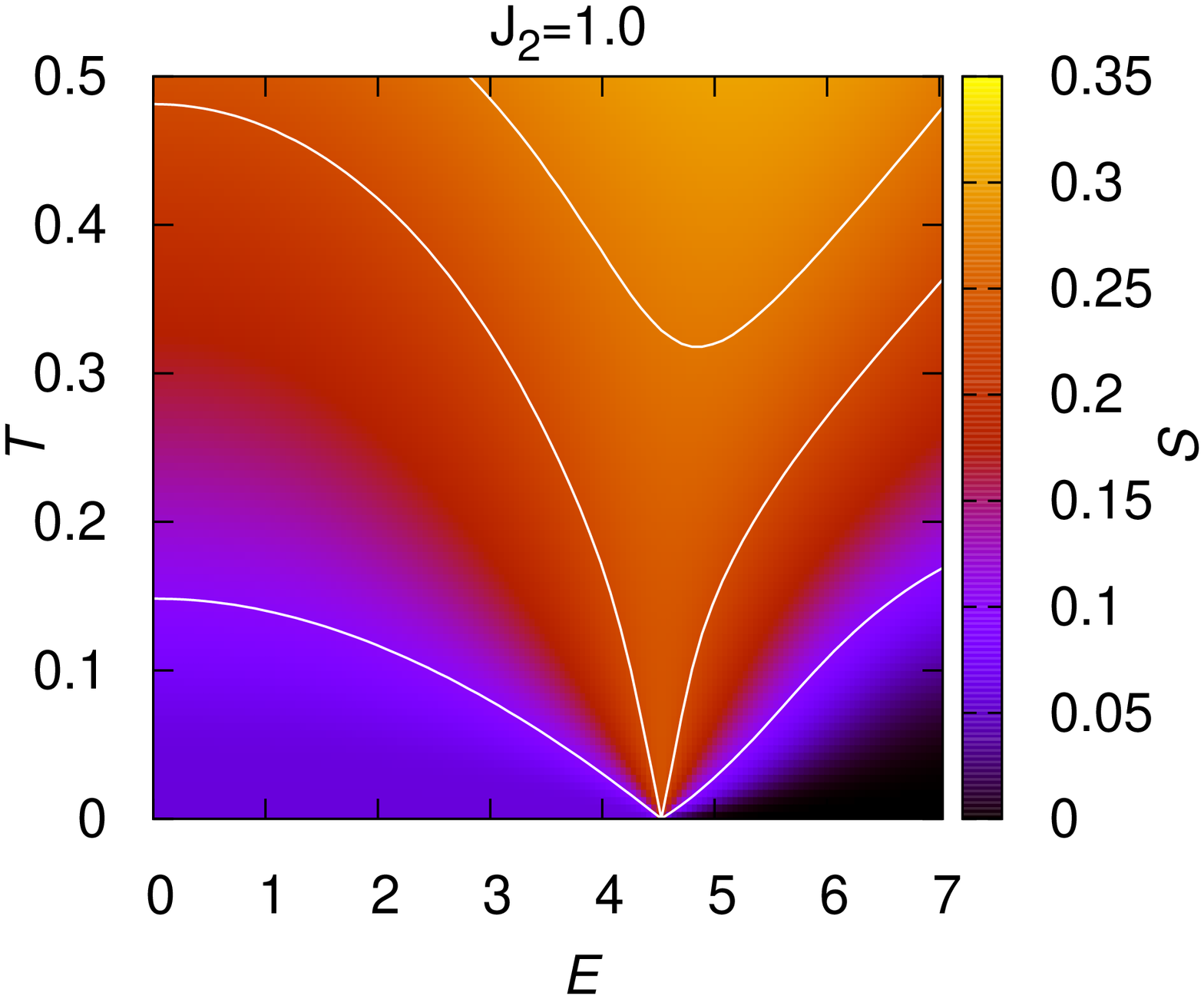}

\includegraphics[clip=on,angle=270,width=50mm]{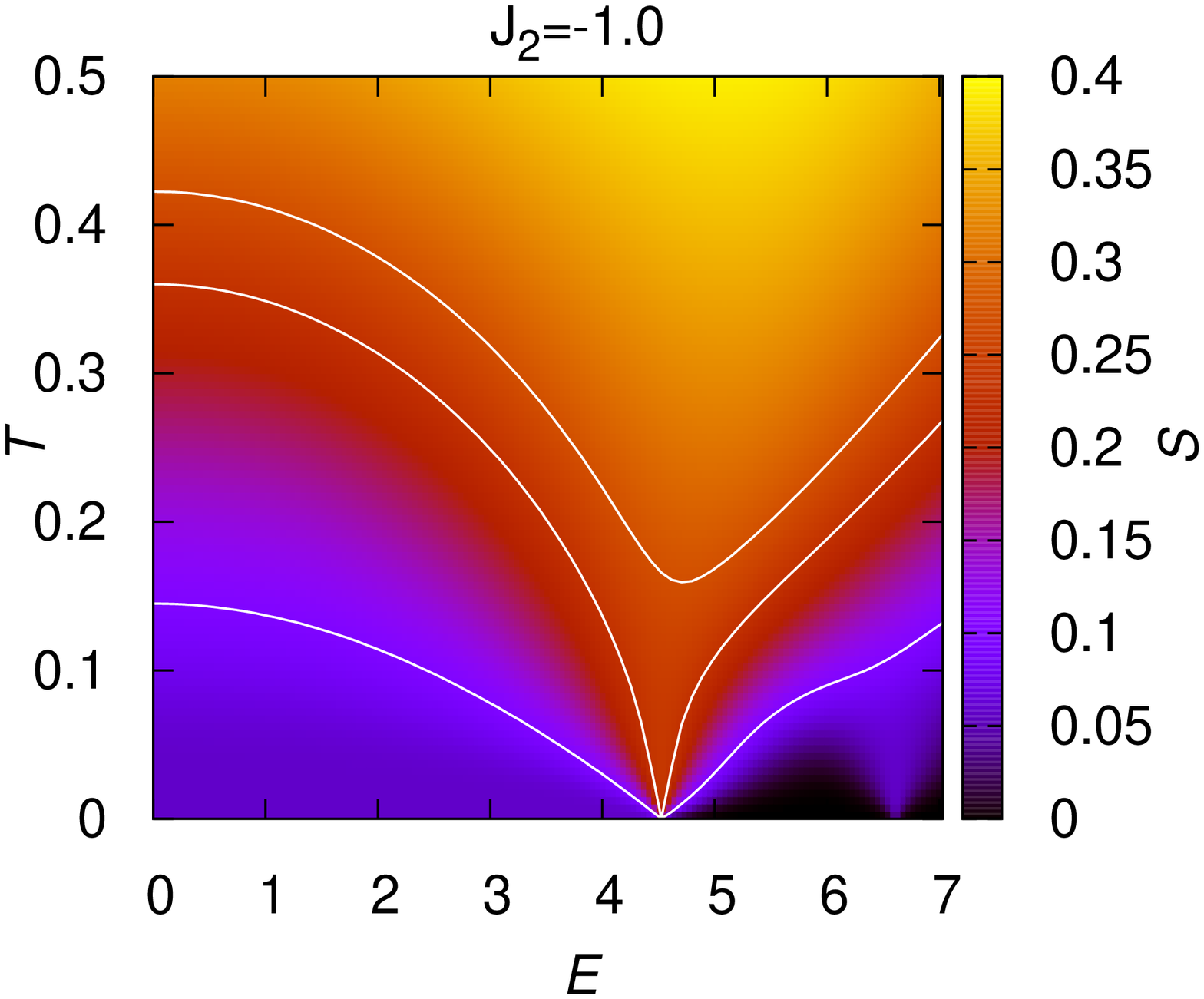}

\includegraphics[clip=on,angle=270,width=50mm]{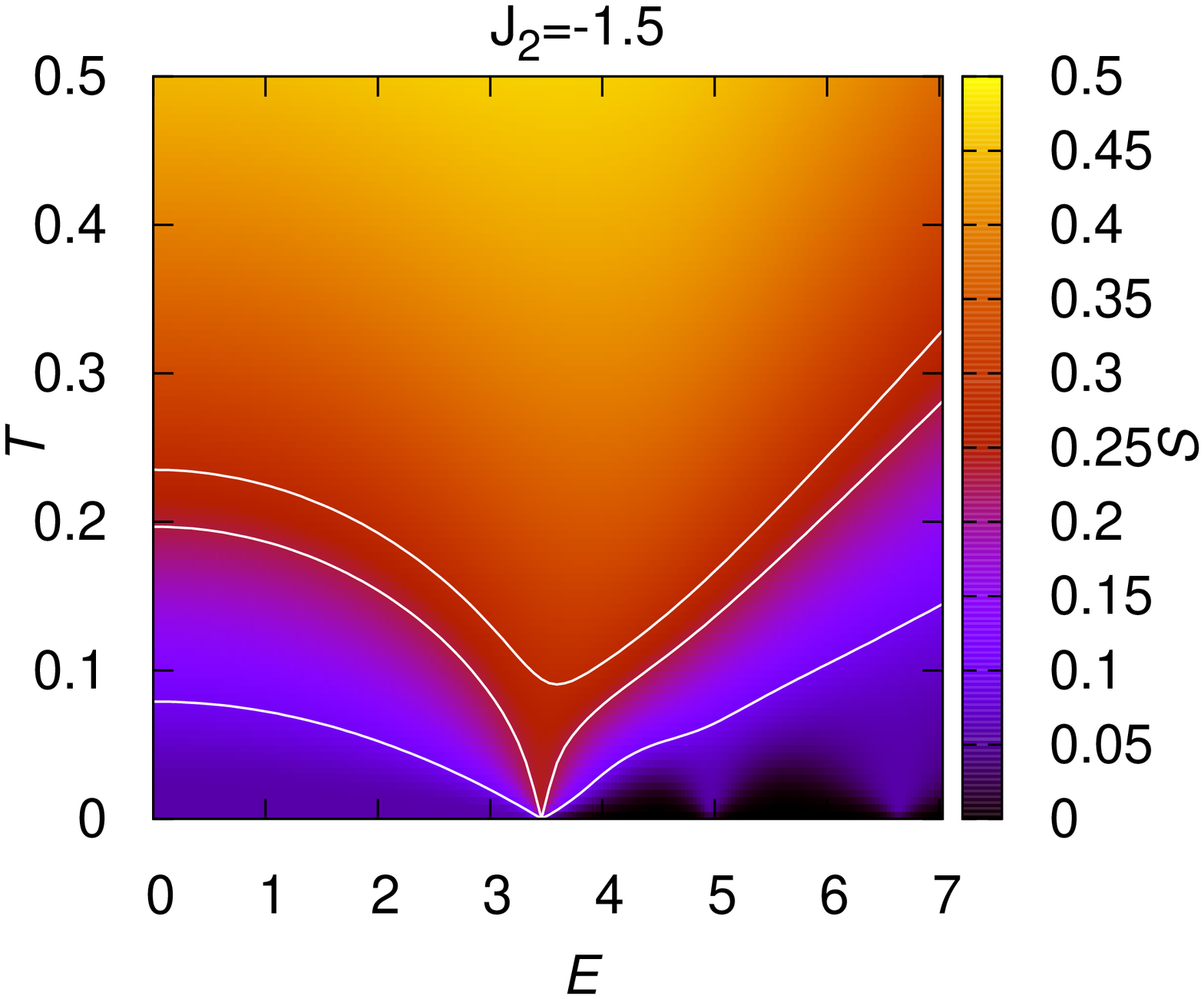}
\caption{(Color online)
Electrocaloric effect (ECE) for the flat-band case I
($\phi=0$,
$\theta=\pi/8$, $B=B_{\rm f}$)
$J_1=1$ and $J_2=1.5$, $1.0$, $-1.0$, $-1.5$.
The white lines represent isentropes for $s=S/N=0.27$, $0.23$ and
$0.1$.  
}
\label{fig6}
\end{center}
\end{figure}

\subsection{Finite-temperature properties }

Again, we focus on flat-band case I.  
As shown by previous studies \cite{lm2,sawt2008,lm5,lm5a,FAF1,FAF3,lm9}
the flat-band states dominate the low-temperature thermodynamics 
at the flat-band point $E=E_{\rm f}$ and $B=B_{\rm f}=B_{\rm sat}$ 
as well as in the vicinity of it. 
This is related to the fact that the    
flat-band (localized multi-magnon) states  
build a 
massively degenerate GS manifold at the  flat-band point, and, 
that in a sizeable parameter region around $E=E_{\rm f}$ and $B=B_{\rm f}$
this former GS manifold acts as 
a huge manifold of low-lying excitations setting an extra low-energy scale.

This is evident by  the temperature profile of the entropy as shown
exemplarily in the main panels of
Fig.~\ref{fig4}, (a) and (b), for $J_1=1$, $J_2=0$ and a few values
of $B$ (resp.\ $E$) at and in the vicinity of the
flat-band value $B_{\rm f}$ (resp.\ $E_{\rm f}$). The pronounced
low-temperature plateaus in $s(T)$ are caused
by the manifold of flat-band states, where the width of these
plateaus depends on the
distance to $B_{\rm f}$ and $E_{\rm f}$, i.e. on $|B-B_{\rm f}|$ and  on
$|E-E_{\rm f}|$, respectively.
\textcolor{black}{The curves for $N=20$, $28$ and $36$ practically coincide down
to very low  temperatures $T$. 
Only below $T \sim 0.005$ small  
finite-size effects appear, i.e. the $s(T)$
profiles shown for various chain length $N$ are slightly different.
The small magnitude of finite-size effects can be attributed to the set of flat-band states
which scale systematically with system size \cite{lm5,lm5a}.} 
Note that for $N \to \infty$ we get $s(T \to 0)= 0$ except for   
the flat-band point, where the system exhibits a non-zero residual entropy.
The different temperature regimes in the $s(T)$ profile are
also obvious in the specific heat $c(T)=T\left({\frac {\partial
S}{\partial T}}\right)$, see the insets in Fig.~\ref{fig4}, (a) and (b).
At $E=E_{\rm f}$, $B=B_{\rm f}$ (red curves in Fig.~\ref{fig4})  
 the huge ground-state manifold  leads to a vanishing specific
heat at low $T$, whereas slightly away  from the flat-band point
the  additional Schottky-like peak in the $c(T)$ profile
indicates the extra low-energy scale. 
It is worth mentioning
that, interestingly, 
for the contribution of the flat-band
states to the partition function 
explicit analytical formulas 
can be found which describe the low-temperature physics near the flat-band
point very well  
\cite{lm2,sawt2008,lm5,lm5a,FAF1,lm9}.

Next, we study the electrocaloric effect (ECE) which currently
attracts enormous attention 
as a promising new approach for refrigeration technologies
\cite{science2006,science2008,science2017,nano2019,science2020}. 
From previous studies we know that due to the large residual entropy
at the flat-band point an enhanced magnetocaloric
effect is observed when traversing the
saturation field (flat-band point) 
\cite{FAF1,cool2004,lm2,lm5,lm5a,63,64,65}.
Consequently, we may expect an extraordinary ECE when pinning the magnetic
field at $B=B_{\rm f}$ and varying the electric field $E$ through the
flat-band value $E_{\rm f}$.
To verify this we study adiabatic cooling, i.e., the isentropic variation of the
temperature when changing
the electric field.
Computationally that is demanding, since an extensive $T-E$ scan is needed to pin the
entropy $s=S/N$ at a predefined value.
Fortunately, again the finite-size effects  are very small as demonstrated
in Fig.~\ref{fig5}. 
Thus we performed extensive simulations of adiabatic
cooling for a small system of $N=12$ to create contour plots of the ECE,
see Fig.~\ref{fig6}.
As already shown in Fig.~\ref{fig5} and in more detail demonstrated by the contour
plots for $J_2=1.5, 1.0, -1.0, -1.5$,  the
variation of the electric field through the
flat-band value $E_{\rm f}$ leads to a strong change of temperature. In
particular, when considering isentropes with $s$ below the residual
entropy $s_{\rm res} \sim 0.24$, the temperature rapidly drops
to zero when approaching  $E=E_{\rm f}$.
Obviously,
the general shape of the isentropes does not depend on $J_2$. However, the
flat-band values $B_{\rm f}$ and  $E_{\rm f}$ depend on $J_2$, see
Eqs.~(\ref{Bsat}) and (\ref{E_1}).
\textcolor{black}{
With respect to a possible realization of such a flat-band induced enhanced  ECE 
in magnetic
compounds 
the question arises whether  
there is still an extraordinary ECE,
even if the ideal flat-band condition is not fulfilled.
For that we show in Fig.~\ref{fig5}(b)  adiabatic cooling by varying the electric field
 when pinning the magnetic
field $B$ at various values below and above the flat-band value $B_{\rm f}$.
Obviously,
the ECE remains enhanced; varying the electric field around $E_{\rm f}$ leads to a
pronounced downturn  of the temperature. 
}

\section{Conclusion}

This paper connects  frustrated quantum magnetism  
(a traditional research field in solid state physics)
and the  Katsura-Nagaosa-Balatsky (KNB) mechanism to couple spin degrees of
freedom to an electric field (a more recent research field in
solid state physics) hereby opening the possibility to study
flat-band effects in quantum magnets by applying an appropriate
electric field.
Prominent effects related to the interplay of frustration and the KNB
mechanism are strongly enhanced 
magnetoelectric and  electrocaloric effects.  

As an example, we consider a specific quantum spin model, the 
spin-half $J_1$-$J_2$ sawtooth Heisenberg chain (see Fig.~\ref{fig1}).
The sawtooth spin chain is a  paradigmatic 
frustrated quantum spin model that can serve as the relevant spin model for
various magnetic compounds
\cite{FAF7,FAF8,FeGd,atacamite,fluoride,euchroite,Mo75V20,Mn18}
such as the recently studied Fe$_{10}$Gd$_{10}$ \cite{FeGd}, Cu$_2$Cl(OH)$_3$ \cite{atacamite}
and  Fe$_2$Se$_2$O$_7$ \cite{sawtooth_exp_2021}.

\textcolor{black}{
Although, we consider a idealized strictly one-dimensional
model we know from experimental studies
\cite{azurite_exp_2005,azurite_exp_2005_b,Tanaka2014}
that the discussed flat-band effects, such as magnetization jumps,
can be seen in real magnetic compounds.
Moreover, several theoretical studies for models violating the flat-band
condition, i.e. models with (slightly) dispersive bands,
demonstrate that flat-band effects are present in  
considerable parameter region around the flat-band point
\cite{Jeschke_azurit2011,azurite_Hon2011,azurite_Derzhko2012,beyond_flat2014,hubbard_nearly_flat2016,richter_sawtooth2020}.}

The large variety of frustrated quantum magnetic insulators
as well as the progress in synthesizing new magnetic molecules and compounds with predefined   
spin lattices may open the window to get access to the observation of the
discussed phenomena.
We expect that the strength of the electric field necessary for that
 can be achieved since typical exchange
couplings are of the order of a few to a hundred kelvin which
corresponds to $1\dots 10$~meV. Depending on the KNB constant
$\gamma$ this could translate into field strengths applicable to
the typically insulating quantum spin materials
where a breakdown voltage of 10-30~MV/m could possibly be achieved.
\textcolor{black}{
  We mention here the very recent experimental study of Kocsis et al. \cite{switsching2021}
for LiCoPO$_4$ in an electric
field of $E \sim 1$ MV/m, where a switching of magnetic states in
LiCoPO$_4$ has been reported.
}

Further we mention that the discussed electric-field driven flat-band physics 
also exists for more general sawtooth-chain models, e.g.\ models
with two different zigzag bonds $J_2$ and $J_3$ including
different XXZ anisotropies on all three bonds. 
Moreover, the flat one-magnon band is also present for
corresponding models with $s>1/2$ \cite{work_in_progress}.
The investigation of the generalized sawtooth model and of other
flat-band spin systems, e.g.\ \cite{lm5,lm5a},
promises a rich future of combined effects of frustration and
KNB mechanism.\\

\section{Acknowledgments}
We thank Hoshu Katsura for illuminating comments on the applicability of the
KNB mechanism on spin systems.
We also thank Paul McClarty and Alexei Andreanov for critical reading the
manuscript.
We thank Hiroyuki Nojiri for
sharing his knowledge on breakdown voltages with us.
V.O. expresses his deep gratitude to the MPIPKS
for warm hospitality during his two-month stay in Dresden.
V. O. also acknowledges the partial financial support form ANSEF (grant
PS-condmatth-2462) and CS RA MESCS (grant 21AG-1C047).
J.R. and J.S. thank the DFG for financial support (grants RI
615/25-1 and SCHN 615/28-1).


\begin{thebibliography}{99}

\bibitem{mee1}
M.~Fiebig,
Journal of Physics D: Applied Physics {\bf 38}, R123 (2005).

\bibitem{Spaldin2008} N. A. Spaldin, M. Fiebig and M. Mostovoy,
J. Phys.: Condens. Matter {\bf 20}, 434203 (2008).

\bibitem{Ederer2008} 
C. Ederer and C. J. Fennie,
J. Phys.: Condens. Matter {\bf 20}, 434219 (2008). 


\bibitem{GBB:S10}
  V. Garcia, M. Bibes, L. Bocher, S. Valencia, F. Kronast,
A. Crassous, X. Moya, S. Enouz-Vedrenne, A. Gloter,
D. Imhoff, C. Deranlot, N. D. Mathur, S. Fusil,
K. Bouzehouane, A. Barth\'el\'emy,
Science  {\bf 327}, 1106 (2010).

\bibitem{mee2}
Y.~Tokura, S.~Seki, and N.~Nagaosa,
Rep. Prog. Phys. {\bf 77}, 076501 (2014).

\bibitem{wang2014} Y. Wang, J. Li, D. Viehland, Mater. Today {\bf 17}, 269
(2014).

\bibitem{dong15} Sh.~Dong, J.-M.~Liu, S.-W.~Cheong, Zh.~Ren,
Adv. Phys. {\bf 64}, 519 (2015).

\bibitem{fie16} M.~Fiebig, T.~Lottermoser, D.~Meier, and M.~Trassin, Nat. Rev. Mater. {\bf 1}, 16046 (2016).

\bibitem{spa19} N.~A.~Spaldin, R.~Ramesh, Nat. Mater. {\bf 18}, 203 (2019).

\bibitem{don19} Sh.~Dong, H.~Xiang, and E.~Dagotto, Nat. Sci. Rev. {\bf 6}, 629 (2019).

\bibitem{app1}
S.-W.~Cheong and M.~Mostovoy,
Nature Materials {\bf 6}, 13 (2007).

\bibitem{app2}
Y.~Tokura and S.~Seki,
Advanced Materials {\bf 22}, 1554 (2010).

\bibitem{LLI:SA21}
J. Liu, V. V. Laguta, K. Inzani, W. Huang, S. Das,
R. Chatterjee, E. Sheridan, S. M. Griffin, A. Ardavan,
R. Ramesh,
Sci. Adv. {\bf 7}, eabf8103 (2021).

\bibitem{MEE_PRL2020}
P. Mellado, A. Concha, and S. Rica,
Phys. Rev. Lett. {\bf  125}, 237602 (2020).

\bibitem{ECE-review}
 J. F. Scott, 
Annu. Rev. Mater. Res. {\bf 41}, 229 (2011).

\bibitem{science2006}
A.S. Mischenko, Q. Zhang, J.F. Scott, R.W. Whatmore, N.D. Mathur,
Science {\bf 311}, 1270 (2006).

\bibitem{science2008} B. Neese, B.J. Chu, S.G. Lu, Y. Wang, E. Furman, Q.M.
 Zhang,
Science {\bf 321}, 821 (2008).

\bibitem{nano2019}
Mengwei Si, Atanu K. Saha, Pai-Ying Liao, Shengjie Gao, Sabine M. Neumayer,
Jie Jian, Jingkai Qin, Nina Balke, Haiyan Wang, Petro Maksymovych, Wenzhuo
Wu, Sumeet K. Gupta, and Peide D. Ye,
ACS Nano {\bf 13}, 8760 (2019).

\bibitem{science2017} 
R. Ma, Z. Zhang, K. Tong, D. Huber, R. Kornbluh, Y. S. Ju, Q. Pei,
 Science {\bf  357}, 1130 (2017).

\bibitem{science2020}
  Y. Wang,  Z. Zhang, T.Usui, M. Benedict, S. Hirose, J. Lee, J. Kalb, D. Schwartz,
 Science {\bf 370}, 129 (2020).



\bibitem{Tsui1982}  
D.C. Tsui, H.L. Stormer, and AS.C. Gossard, 
Phys. Rev. Lett. {\bf 48},  1559 (1982).


\bibitem{25} H. Tasaki,
Phys. Rev. Lett., {\bf 69}, 1608 (1992).

\bibitem{26} A. Mielke, H. Tasaki,
Commun. Math. Phys.,  {\bf 158}, 341 (1993).

\bibitem{27} H. Tasaki,
      J. Stat. Phys., {\bf  84}, 535 (1996).

\bibitem{2} J. Schulenburg, A. Honecker, J. Schnack, J. Richter, and
      H.-J. Schmidt,
      Phys. Rev. Lett. {\bf 88}, 167207 (2002).

\bibitem{spin-peierls} J. Richter, O. Derzhko, and J. Schulenburg,
Phys. Rev. Lett. {\bf 93}, 107206 (2004).

\bibitem{lm2} M.~E.~Zhitomirsky and H.~Tsunetsugu, Phys. Rev. B {\bf 70},
100403(R)
(2004).

\bibitem{huber2010}
S. D. Huber and E. Altman,
Phys. Rev. B {\bf 82}, 184502 (2010).

\bibitem{kastura2010}
H. Katsura, I. Maruyama, A. Tanaka, and H. Tasaki,
EPL {\bf 91}, 57007 (2010).
\bibitem{kastura2011}
K. Sun, Z. Gu, H. Katsura, and S. DasSarma,
Phys. Rev. Lett. {\bf 106} 236803 (2011).
\bibitem{JGT:PRL12}
Gyu-Boong Jo, J. Guzman, C.K. Thomas, 
P. Hosur, A. Vishwanath, and D.M. Stamper-Kurn,
Phys. Rev. Lett. {\bf 108}, 045305 (2012).

\bibitem{SOW:PRL12}
 J. Struck,  C. \"Olschl\"ager, M. Weinberg, P. Hauke, 
 J. Simonet, A. Eckardt, M. Lewenstein, K. Sengstock, and
 P. Windpassinger,
Phys. Rev. Lett. {\bf 108}, 225304
 (2012).

\bibitem{bergholtz2013}
E.J. Bergholtz and Zhao Liu,
Int. J. Mod. Phys. B {\bf 27}, 1330017 (2013).
\bibitem{sondhi2013}
S. A. Parameswaran, R. Roy, and S. L. Sondhi,
C. R. Phys. {\bf 14}, 816
(2013).
\bibitem{leykam2013}
D. Leykam, S. Flach, O. Bahat-Treidel, and A. S. Desyatnikov,
Phys. Rev. B {\bf 88}, 224203 (2013).

\bibitem{lm6} O.~Derzhko, J.~Richter, and M.~Maksymenko,
Int. J. Mod. Phys.
{\bf 29}, 1530007 (2015).

\bibitem{VCM:PRL15}
R.A. Vicencio, C. Cantillano,
L. Morales-Inostroza, B. Real, C. Mej\'{\i}a-Cort\'es,
S. Weimann, A.Szameit, and M.I. Molina, 
Phys. Rev. Lett. {\bf 114}, 245503 (2015).

\bibitem{MSC:PRL15}
S. Mukherjee, A. Spracklen, D. Choudhury,
N. Goldman, P. \"Ohberg, E. Andersson, and
R.R. Thomson,
Phys. Rev. Lett. {\bf 114}, 245504 (2015).

\bibitem{graphene2018}
    Y. Cao, V. Fatemi, S. Fang, K. Watanabe, T. Taniguchi,
 E. Kaxiras, and P. Jarillo-Herrero, 
Nature {\bf 556}, 43 (2018).

\bibitem{leykam2018} D.~Leykam, A.~Andreanov, and S.~Flach, Adv. Phys.: X {\bf 3}, 1473052 (2018).

\bibitem{photonic} L. Tang, D. Song, S. Xia, Shiqiang Xia, J.
     Ma, W. Yan, Yi Hu, J. Xu, D. Leykam, and Z. Chen,
Nanophotonics {\bf  9}, 1161 (2020).


\bibitem{graphene2020a}
    M. Serlin, C. L. Tschirhart, H. Polshyn, Y. Zhang, J. Zhu, 
K. Watanabe, T. Taniguchi, L. Balents, A.F. Young,
Science {\bf 367}, 900 (2020).
\bibitem{graphene2020b}
    L. Balents, C. R. Dean, D. K. Efetov, and A. F. Young 
Nat. Phys. {\bf 16}, 725 (2020).
\bibitem{graphene2020c}
    P. Stepanov, I. Das, X. Lu, A. Fahimniya, K. Watanabe,
 T. Taniguchi, F. H. L. Koppens, J. Lischner, L. Levitov,
 and
 D. K. Efetov 
Nature {\bf 583}, 375 (2020).


\bibitem{mag_cryst_exp}
R. Okuma, D. Nakamura, T. Okubo, A. Miyake, A. Matsuo, K. Kindo, M. Tokunaga, 
N. Kawashima, S. Takeyama, and Z. Hiroi,
Nat. Commun. {\bf 10}, 1229 (2019).


\bibitem{prl2020}
 J. Schnack, J. Schulenburg, A. Honecker, and J. Richter
Phys. Rev. Lett. {\bf 125}, 117207 (2020).


\bibitem{KNB1}
H.~Katsura, N.~Nagaosa, and A.~V.~Balatsky,
Phys. Rev. Lett. {\bf 95}, 057205 (2005).

\bibitem{KNB3}
\textcolor{black}{H.~Katsura,
 Nevill F. Mott Prize
Lecture at the SCES 2016, Hangshou (May 2016).}

\bibitem{KNB2}
C.~Jia, S.~Onoda, N.~Nagaosa, and J.~H.~Han,
Phys. Rev. B {\bf 74}, 224444 (2006).


\bibitem{bro13}
M.~Brockmann, A.~Kl\"{u}mper, and V.~Ohanyan,
Phys. Rev. B {\bf 87}, 054407 (2013).

\bibitem{Berakdar2014} 
M. Azimi, L. Chotorlishvili, S. K. Mishra, S. Greschner, T. Vekua,
and J. Berakdar, Phys. Rev. B {\bf 89},  024424 (2014).

\bibitem{Berakdar2016} M. Azimi,  M. Sekania,  S. K. Mishra,  L. Chotorlishvili,  Z. Toklikishvili, and J.
Berakdar,
Phys. Rev. B {\bf 94},  064423  (2016).

\bibitem{Berakdar2017} 
S. Stagraczynski,  L. Chotorlishvili,  M. Sch\"uler,  M. Mierzejewski,  and J. Berakdar, 
Phys. Rev. B {\bf 96}, 054440 (2017).


\bibitem{thakur18} P.~Thakur, P.~Durganandini, Phys. Rev. B {\bf 97}, 064413 (2018).

\bibitem{XYZ} T.-Ch.~Yi, W.-L.~You, N. Wu, and A.~M.~Ole\'{s}, Phys. Rev. B {\bf 100}, 024423 (2019).


\bibitem{White2020} 
N. Reynolds, A. Mannig, H. Luetkens, C. Baines, T. Goko, R. Scheuermann, L. Keller, M. Bartkowiak, A. Fujimura, Y. Yasui, Ch. Niedermayer, and J. S.
White,
Phys. Rev. B {\bf 99}, 214443 (2019).

\bibitem{oha20} 
V.~Ohanyan, Condens. Matter Phys. {\bf 23}, 43704  (2020).

\bibitem{mench15} O.~Menchyshyn, V.~Ohanyan, T.~Verkholyak, T.~Krokhmalskii, and O.~Derzhko,
Phys. Rev. B {\bf 92}, 184427 (2015).

\bibitem{sznajd18} J.~Sznajd, Phys. Rev. B {\bf 97}, 214410 (2018).

\bibitem{sznajd19} J.~Sznajd, J. Magn. Magn. Mater. {\bf 479}, 254 (2019).

\bibitem{baran18} O.~Baran, V.~Ohanyan, and T.~Verkholyak, Phys. Rev. B {\bf 98}, 064415 (2018).

\bibitem{oles}
W.-L.~You, G.-H.~Liu, P.~Horsch, and A.~M.~Ole\'{s},
Phys. Rev. B {\bf 90}, 094413 (2014).


\bibitem{stre20} J.~Stre\v{c}ka, L. G\'{a}lisov\'{a}, and T.~Verkholyak, Phys. Rev. E {\bf 101}, 012103 (2020).

\bibitem{Enderle2005} M.~Enderle, C.~Mukherjee, B.~Fak, R.K.~Kremer,
J.-M.~Broto,
        H.~Rosner, S.-L.~Drechsler, J.~Richter, J.~Malek, A.~Prokofiev,
        W.~Assmus, S.~Pujol, J.-L.~Raggazoni, H.~Rakato, M.~Rheinst\"adter,
        and H.M.~Ronnow,
        EPL {\bf 70}, 237 (2005).

\bibitem{KBK:PRB06}
        R.~Klingeler, B. B{\"u}chner, K.-Y. Choi,
  V. Kataev, U. Ammerahl, A. Revcolevschi, J. Schnack,
Phys. Rev. B {\bf 73}, 014426 (2006).

\bibitem{LiCu2O21}
S.~Park, Y.~J.~Choi, C.~L.~Zhang, and S.-W.~Cheong,
Phys. Rev. Lett. {\bf 98}, 057601 (2007).


\bibitem{Drechsler2007} S.-L. Drechsler,   O.\ Volkova,  A.N.\ Vasiliev, N.~Tristan,  J.\ Richter,
        M. Schmitt,
        H. Rosner, J.\ M\'alek, R.\ Klingeler, A.A.~Zvyagin, and B.\
        B\"uchner,
        Phys. Rev. Lett. {\bf 98}, 077202 (2007).

\bibitem{LiVCuO41}
Y. Yasui, Y. Naito, K. Sato, T. Moyoshi, M.
Sato, and K. Kakurai, 
J. Phys. Soci. Japan {\bf 77}, 023712 (2008).

\bibitem{LiVCuO42}
F. Schrettle, S. Krohns, P. Lunkenheimer, J. Hemberger, N. B\"uttgen,
H.-A. Krug von Nidda, A. V. Prokofiev, and A. Loidl, 
Phys. Rev. B {\bf 77}, 144101 (2008).


\bibitem{LiCuO21}
S.~Seki, Y.~Yamasaki, M.~Soda, M.~Matsuura, K.~Hirota, and Y.~Tokura,
Phys. Rev. Lett. {\bf 100}, 127201 (2008).

\bibitem{LiCu2O22}
A.~S.~Moskvin, Yu.~D.~Panov, S.-L.~Drechsler,
Phys. Rev. B {\bf 79}, 104112 (2009).






\bibitem{LiCuVO43} A.~S.~Moskvin, S.-L. Drechsler,
EPL {\bf 81}, 57004 (2008).

\bibitem{LiCuVO44} Y.~Yasui, K.~Sato, Y.~Kobayashi, and M.~Sato, J. Phys. Soc. Jpn. {\bf 78}, 084720 (2009).

\bibitem{LiCuO23}
Y.~Qi and A.~Du,
Phys. Lett. A {\bf 378}, 1417 (2014).

\bibitem{lm5}  O.~Derzhko, J.~Richter,
Eur. Phys. J. B {\bf 52}, 23 (2006).

\bibitem{lm5a}
O. Derzhko, J. Richter, A. Honecker, and H.-J. Schmidt,
Low Temp. Phys. {\bf 33}, 745 (2007).

\bibitem{lm4} J.~Richter, J.~Schulenburg, A.~Honecker, J.~Schnack, and H.~J.~Schmidt,
 J. Phys.: Condens. Matter {\bf 16}, S779 (2004).


\bibitem{lm2a} M. E. Zhitomirsky and H. Tsunetsugu,
Prog. Theor. Phys. Suppl. {\bf 160}, 361 (2005).

\bibitem{lm3} O.~Derzhko and J.~Richter, Phys. Rev. B {\bf 70}, 104415 (2004).

\bibitem{cool2004} M. E. Zhitomirsky and A. Honecker,
J. Stat. Mech.,  P07012 (2004).

\bibitem{sawt2008} J.~Richter, O.~Derzhko, A.~Honecker,
       Int. J. Modern Phys. B {\bf 22}, 4418 (2008). 

\bibitem{FAF1} V.~Ya.~Krivnov, D.~V.~Dmitriev, S.~Nishimoto, S.-L.~Drechsler,
and J.~Richter, Phys. Rev. B {\bf 90}, 014441 (2014).

\bibitem{lm8} A.~Metavitsiadis, C.~Psaroudaki, and W.~Brenig, Phys. Rev. B {\bf 101}, 235143 (2020).



\bibitem{lm9} O.~Derzhko, J.~Schnack, D.~V.~Dmitriev, V.~Ya.~Krivnov, J.~Richter, Eur. Phys. J. B {\bf 93}, 161 (2020).

\bibitem{dre20}
T.~Yamaguchi ,  S.-L.~Drechsler, Y.~Ohta ,  and S.~Nishimoto,
Phys. Rev. B {\bf 101}, 104407 (2020).

\bibitem{McClarty2020} P. A. McClarty, M. Haque, A. Sen, and J. Richter,
Phys. Rev. B  {\bf 102}, 224303 (2020).

\bibitem{Hatsugai2020}
Y. Kuno, T. Mizoguchi, and Y. Hatsugai
Phys. Rev. B {\bf 102}, 241115(R) (2020).


\bibitem{Pujol2020}
S. Acevedo, P. Pujol, C.A. Lamas
Phys. Rev. B {\bf 102}, 195139 (2020).


\bibitem{FAF2} D.~V.~Dmitriev and V.~Ya.~Krivnov, J. Phys.: Condens. Matter
{\bf 28}, 506002 (2016).



\bibitem{28} Y. Watanabe and S. Miyashita,
J. Phys. Soc. Jpn., {\bf  66}, 2123 (1997).


\bibitem{29} R. Arita, Y. Shimoi, K. Kuroki, and H. Aoki,
Phys. Rev. B, {\bf 57}, 10609 (1998).

\bibitem{hubb2010}
O. Derzhko, J. Richter, A. Honecker, M. Maksymenko, and  R.
      Moessner,
      Phys. Rev. B {\bf 81}, 014421 (2010).


\bibitem{31} M. Maksymenko, A. Honecker, R. Moessner, J. Richter, and O. Derzhko,
Phys. Rev. Lett.,  {\bf 109}, 096404 (2012).


\bibitem{32} S. Flach, D. Leykam, J. D. Bodyfelt, P.
     Matthies, and A. S. Desyatnikov,
EPL, {\bf  105}, 30001 (2014).

\bibitem{maimaiti2017}
W. Maimaiti, A. Andreanov, H. C. Park, O. Gendelman, and S. Flach,
Phys. Rev. B {\bf 95}, 115135 (2017).  

\bibitem{33} W. Zhang, R. Liu, and W. Nie,
     Science Bulletin, {\bf 64}, 1490 (2019).

\bibitem{Katsura_delta1}
K. Tamura and H. Katsura,
Phys. Rev. B {\bf 100}, 214423  (2019).

\bibitem{Katsura_delta2}
K. Tamura and H. Katsura,
J. Stat. Phys. {\bf 182}, 16 (2021).





\bibitem{34} S. Weimann, L. Morales-Inostroza, B. Real,
     C. Cantillano, A. Szameit, and R. A. Vicencio,
Opt. Lett. {\bf  41}, 2414 (2016).

\bibitem{lin_indep}
H.-J. Schmidt, J. Richter, R. Moessner,
       J. Phys. A.: Math. Gen. {\bf 39}, 10673 (2006).



\bibitem{Tanaka2014}
H.~Tanaka, N.~Kurita, M.~Okada, E.~Kunihiro, Y.~Shirata, K.~Fujii,
H.~Uekusa, A.~Matsuo, K.~Kindo, and H.~Nojiri,
J. Phys. Soc. Jpn. {\bf 83}, 103701 (2014).

\bibitem{bil2018}
J. Richter, O. Krupnitska, V. Baliha, T. Krokhmalskii, and
O. Derzhko,
     Phys. Rev. B {\bf 97}, 024405 (2018).



\bibitem{FAF4} T.~Tonegawa and M.~Kaburagi, J. Magn. Magn. Mater. {\bf  272-276},
898 (2004).

\bibitem{FAF5} M.~Kaburagi, T.~Tonegawa, and M.~Kang, J. Appl. Phys. {\bf 97},
10B306 (2005).

\bibitem{FAF3} D.~V.~Dmitriev, V.~Ya.~Krivnov,  J.~Richter, and J.~Schnack, Phys. Rev. B {\bf 99}, 094410  (2019).

\bibitem{FAF_field} D.~V.~Dmitriev, V.~Ya.~Krivnov, J.~Schnack, and J.~Richter, Phys. Rev. B {\bf 101}, 054427 (2020).

\bibitem{FAF7} C.~Ruiz-Perez, M.~Hernandez-Molina, P.~Lorenzo-Luis, F.~Lloret, J.~Cano, and M.~Julve,
 Inorg. Chem. {\bf 39}, 3845 (2000).
\bibitem{FAF8}
Yuji Inagaki, Yasuo Narumi, Koichi Kindo, Hikomitsu Kikuchi, Tomohisa
Kamikawa, Takashi Kunimoto,
Susumu Okubo, Hitoshi Ohta, Takashi Saito, Masaki Azuma, Mikio
Takano, Hiroyuki Nojiri, Makoto Kaburagi, and Takashi Tonegawa,
J. Phys. Soc. Jpn. {\bf 74}, 2831 (2005).

\bibitem{FeGd} A.~Baniodeh, N.~Magnani, Y.~Lan, G.~Buth, C.~E.~Anson,
J.~Richter, M.~Affronte, J.~Schnack, and A.~K.~Powell, npj Quantum Mater. {\bf 3}, 10 (2018).


\bibitem{atacamite}
L. Heinze, H. O. Jeschke, I. I. Mazin, A. Metavitsiadis, M. Reehuis, R. Feyerherm, 
J.-U. Hoffmann, M. Bartkowiak, O. Prokhnenko, A. U. B. Wolter, X. Ding, 
V. S. Zapf, C. Corvalan Moya, F. Weickert, M. Jaime, K. C. Rule, 
D. Menzel, R. Valenti, W. Brenig, S. S\"ullow,
Phys. Rev. Lett. {\bf 126}, 207201 (2021).

 \bibitem{fluoride}
R. Shirakami, H. Ueda, H. O. Jeschke, H. Nakano, S.
Kobayashi, A. Matsuo,
T. Sakai, N. Katayama, H. Sawa, K. Kindo, C. Michioka,
and K. Yoshimura,
Phys. Rev. B {\bf 100}, 174401 (2019).

\bibitem{euchroite}
H. Kikuchi, Y. Fujii, D. Takahashi, M. Azuma, Y. 
Shimakawa, T. Taniguchi, A. Matsuo and K. Kindo,
J. Phys. Conf. Series {\bf 320}, 12045 (2011).

\bibitem{Mo75V20}
Y. Oshima, H. Nojiri, J. Schnack, P. K\"ogerler,
M. Luban,
Phys. Rev. B {\bf 85}, 024413 (2012).


\bibitem{Mn18}
M. Coletta, T. G. Tziotzi, M. Gray, G. S. Nichol, M. K.
Singh, C. J. Milios
and E. K. Brechin,
Chemical Communications (2021)
DOI:10.1039/D1CC00185J.

\bibitem{sawtooth_exp_2021}
\textcolor{black}{ 
K. Nawa, M. Avdeev, P. Berdonosov, A. Sobolev 
I. Presniakov, A. Aslandukova, E. Kozlyakova, A. Vasiliev, I. Shchetinin, and T. J. Sato,
Sci. Rep. {\bf 11}, 24049 (2021).}


\bibitem{49} J. Schulenburg, spinpack-2.59 , Magdeburg University
      (2019).
\bibitem{50} J. Richter and J. Schulenburg, 
Eur. Phys. J. B {\bf  73}, 117 (2010).


\bibitem{Accuracy2020}  J. Schnack, J. Richter, and R. Steinigeweg,
   Phys. Rev. Research {\bf 2}, 013186 (2020).

\bibitem{51} J. Jaklic and P. Prelovsek, 
Phys. Rev. B {\bf   49}, 5065 (1994).
\bibitem{52} J. Jaklic and P. Prelovsek, 
Adv. Phys. {\bf 49}, 1 (2000).
\bibitem{53} A. Hams and H. De Raedt, 
Phys. Rev. E {\bf  62}, 4365 (2000).
\bibitem{54} J. Schnack and O. Wendland, 
Eur. Phys. J. B {\bf 78}, 535 (2010).
\bibitem{55} 
P. Prelovsek and J. Bonca, "Ground State and Finite Temperature {Lanczos} Methods",
in
"Strongly Correlated Systems, Numerical Methods",
eds.
A. Avella and F. Mancini,
Springer Series in Solid-State Sciences 176
(Berlin, 2013)

\bibitem{56} O. Hanebaum and J. Schnack, 
Eur. Phys. J. B {\bf   87}, 194 (2014).
\bibitem{57} B. Schmidt and P. Thalmeier, 
Phys. Rep. {\bf  703}, 1 (2017).
\bibitem{58} E. Pavarini, E. Koch, R. Scalettar, and
      R. M. Martin, eds., 'The physics of correlated insulators, metals, and superconductors',  (2017) Chap.
      'The Finite Temperature Lanczos Method and its Applications' by P. Prelovsek, ISBN 978-3-95806-224-5,
      http://hdl.handle.net/2128/15283.
\bibitem{59} J. Schnack, J. Schulenburg, and J. Richter, 
Phys. Rev. B {\bf  98}, 094423 (2018).
\bibitem{61} K. Seki and S. Yunoki, 
Phys. Rev. B {\bf  101}, 235115 (2020).
\bibitem{62}
K. Morita and T. Tohyama,  
 Phys. Rev. Research {\bf 2}, 013205 (2020).
\bibitem{63} M. E. Zhitomirsky, 
Phys. Rev. B {\bf  67}, 104421 (2003).

\bibitem{64} J. Schnack, R. Schmidt, and J. Richter,
 Phys. Rev. B {\bf   76}, 054413 (2007).
\bibitem{65} B. Wolf, A. Honecker, W. Hofstetter, U.
        Tutsch, and M. Lang, 
        Int. J.  Mod. Phys. B {\bf  28}, 1430017 (2014).

\bibitem{azurite_exp_2005}
\textcolor{black}{H. Kikuchi, Y. Fujii, M. Chiba, S. Mitsudo, T. Idehara, T. Tonegawa, K. Okamoto, T. Sakai, T. Kuwai, and H.
Ohta,
Phys. Rev. Let. B {\bf 94}, 227201 (2005).}

\bibitem{azurite_exp_2005_b}
\textcolor{black}{H. Kikuchi, Y. Fujii, M. Chiba, S. Mitsudo, T. Idehara, T. Tonegawa, K. Okamoto,
T. Sakai, T. Kuwai, K. Kindo, A. Matsuo, W. Higemoto,
K. Nishiyama,  M. Horvatic  and C. Berthier,
Prog. Theor. Phys. Suppl. {\bf 159}, 1 (2005).}




\bibitem{Jeschke_azurit2011}  
\textcolor{black}{
H. Jeschke,  I. Opahle, H. Kandpal, R. Valenti, H. Das, T.
Saha-Dasgupta,
       O. Janson, H. Rosner, A. Bruhl, B. Wolf, M. Lang, J. Richter,
       S. Hu, X. Wang, R. Peters,  T. Pruschke, and A. Honecker,      
Phys. Rev. Lett. {\bf 106}, 217201 (2011).
}

\bibitem{azurite_Hon2011} \textcolor{black}{  A. Honecker, S. Hu, R. Peters, and J. Richter,
       J. Phys.: Condens. Matter {\bf 23} 164211 (2011).
}

\bibitem{azurite_Derzhko2012} \textcolor{black}{ O. Derzhko, J. Richter, O. Krupnitska,
       Cond. Matter Phys. {\bf 15}, 43702 (2012).
}

\bibitem{beyond_flat2014}
\textcolor{black}{
 O.~Derzhko, J. Richter, O.~Krupnitska, and T.
Krokhmalski,
        Low Temp. Phys. {\bf 40}, 662 (2014).
}


\bibitem{hubbard_nearly_flat2016}  
\textcolor{black}{
P. M\"uller, J. Richter, and O. Derzhko,
Phys. Rev. B {\bf  93}, 144418 (2016). 
}


\bibitem{richter_sawtooth2020} 
\textcolor{black}{
J. Richter, J. Schulenburg, D.V. Dmitriev, V.Ya. Krivnov,
      and J. Schnack,
   Cond. Matter Phys. {\bf 23}, 43710  (2020).
}



\bibitem{switsching2021}
\textcolor{black}{ V. Kocsis, Y. Tokunaga, Y. Tokura, and Y. Taguchi, 
 Phys. Rev. B {\bf 104}, 054426 (2021).}

 \bibitem{work_in_progress} V. Ohanyan, J. Richter, and A.Andreanov, work in progress.




\end{thebibliography}
\end{document}